\documentclass[notitlepage, twocolumn,superscriptaddress,showkeys]{revtex4-1}

\usepackage[T1]{fontenc}
\usepackage{graphicx}

\usepackage{amsmath}
\usepackage{times}
\usepackage[version=4]{mhchem}
\usepackage{array}

\newcommand{\etal}{\textit{et al}.\ }

\usepackage{xcolor}

\begin{document}

\title{Screening of material defects using universal machine-learning interatomic potentials}
\author{Ethan Berger}
\affiliation{Microelectronics Research Unit, Faculty of Information Technology and Electrical Engineering, University of Oulu, P.O. Box 4500, Oulu, FIN-90014, Finland}
\affiliation{Department of Physics, Chalmers University of Technology, SE-41296 Gothenburg, Sweden}
\author{Mohammad Bagheri}
\affiliation{Nanoscience Center, Department of Physics, University of Jyv\"askyl\"a, Finland}
\author{Hannu-Pekka Komsa}
\email{hannu-pekka.komsa@oulu.fi}
\affiliation{Microelectronics Research Unit, Faculty of Information Technology and Electrical Engineering, University of Oulu, P.O. Box 4500, Oulu, FIN-90014, Finland}

\begin{abstract}
Finding new materials with previously unknown atomic structure or materials with optimal set of properties for a specific application greatly benefits from computational modeling. Recently, such screening has been dramatically accelerated by the invent of universal machine-learning interatomic potentials that offer first principles accuracy at orders of magnitude lower computational cost. Their application to the screening of defects with desired properties or to finding new stable compounds with high density of defects, however, has not been explored. Here, we show that the universal machine-learning interatomic potentials have reached sufficient accuracy to enable large-scale screening of defective materials. We carried out vacancy calculations for 86 259 materials in the Materials Project database and analyzed the formation energies in terms of oxidation numbers.
We further demonstrate the application of these models for finding new materials at or below the convex hull of known materials and for simulated etching of low-dimensional materials.
\end{abstract}

\keywords{defects, vacancies, machine-learning interatomic potential, benchmark, 2D materials}

\maketitle

\section{Introduction}

Computer simulations of materials based on first principles methods, quantum chemistry and density-functional theory, are playing an increasingly important role in the discovery of new materials. On one hand, materials with optimal properties for specific applications can be found by carrying out explicit high-throughput calculations or employing databases of previously calculated data \cite{Curtarolo13_NMat,Alberi19}. These predictions can then be experimentally verified, as demonstrated in the case of magnetic materials \cite{Sanvito17_SciAdv}, high-entropy alloys \cite{Divilov24_Nat}, thermoelectrics \cite{Deng24_AM}, and dielectrics \cite{Riebesell2024}, to name a few. On the other hand, new stable materials, i.e., falling below the convex hull of all previously known stable compounds, can be discovered by atom substitution \cite{Wang21_npjCM}, crystal structure prediction \cite{Oganov19_NRM}, simulated etching \cite{Bjork2024}, or random structure search \cite{Merchant23gnome,Zeni25mattergen}. 

In recent years, the progress in materials research has been massively accelerated by the development of machine-learning (ML) methods and in the context of atomic simulations machine-learning interatomic potentials (MLIPs), which can yield results nearly matching the accuracy of first-principles calculations at a fraction of the computational cost \cite{Deringer19_AM,Wang24_iS}.
Going further, to eliminate the burden of training application-specific MLIPs, universal MLIPs (UMLIP) have been developed with the intention to work for any system with any elements.
These are trained on the massive datasets collected from Materials Project (MP) \cite{Jain2013_MP}, Alexandria \cite{Schmidt23_AM}, or Open catalyst project \cite{Tran23oc}.
Over the last three years only, many competing UMLIPs have been reported, such as 
M3GNet \cite{Chen2022_m3gnet}, CHGNet \cite{Deng_2023_chgnet}, ALIGNN \cite{Choudhary2023_alignn}, MACE \cite{Batatia2022mace,Batatia2023foundation}, GNoME \cite{Merchant23gnome}, MatterSim \cite{Yang24mattersim}, EquiformerV2 \cite{Liao24eqv2}, ORB \cite{Neumann24orb}, and SevenNet \cite{Park24sevennet}.
The leading universal potentials are showing few tens of meV accuracy on formation energies of stable compounds \cite{Riebesell2024matbench}, which is comparable to the energy differences observed near the convex hull and thus the modern UMLIPs are highly promising for accelerated screening of new stable materials.

It is then somewhat surprising that the applicability of UMLIPs for defect screening has not been properly benchmarked to date. This may be due to the fact that there does not exist a similar large dataset of defect calculations, although several smaller databases targeting a set of materials or a set of defects have been collected \cite{Angsten_2014,Huang2023,Bjork2024,Davidsson2023_db,Rahman2024,Kumagai21_PRM,Witman2023,Ivanov2023}. The training sets for UMLIPs do not explicitly contain defective materials, although motifs resembling their atomic structures might still be present in the training set. Nevertheless, this is expected to lead to lower accuracy for defective systems. Fortunately, screening defects is also more forgiving as they often exhibit a large range of values up to 10 eV. For finding the defects with lowest formation energy, determining whether they are likely to be present in high concentration under given conditions (given a choice of chemical potentials), or if the formation energies are negative which entails material decomposition, accuracy of hundreds of meVs can be sufficient \cite{Freysoldt14_RMP,Ibragimova22_CM,Bjork2023}.

In this paper, we start by benchmarking UMLIPs for defect calculations by collating results from several existing targeted defect databases. As the UMLIPs are found to exhibit accuracy sufficient for defect screening, we carry out further analysis. We first calculate vacancy formation energies for all materials in the Materials Project database and analyze the trends in terms of oxidation numbers. We then demonstrate how these calculations can be used to find new materials near the convex hull, either stable compounds or materials that are likely to host large concentration of vacancies. Finally, we discover new two-dimensional materials via simulated non-equilibrium etching from stable parent phases.

\section{Results}

\begin{figure*}
    \centering
    \includegraphics[width=\linewidth]{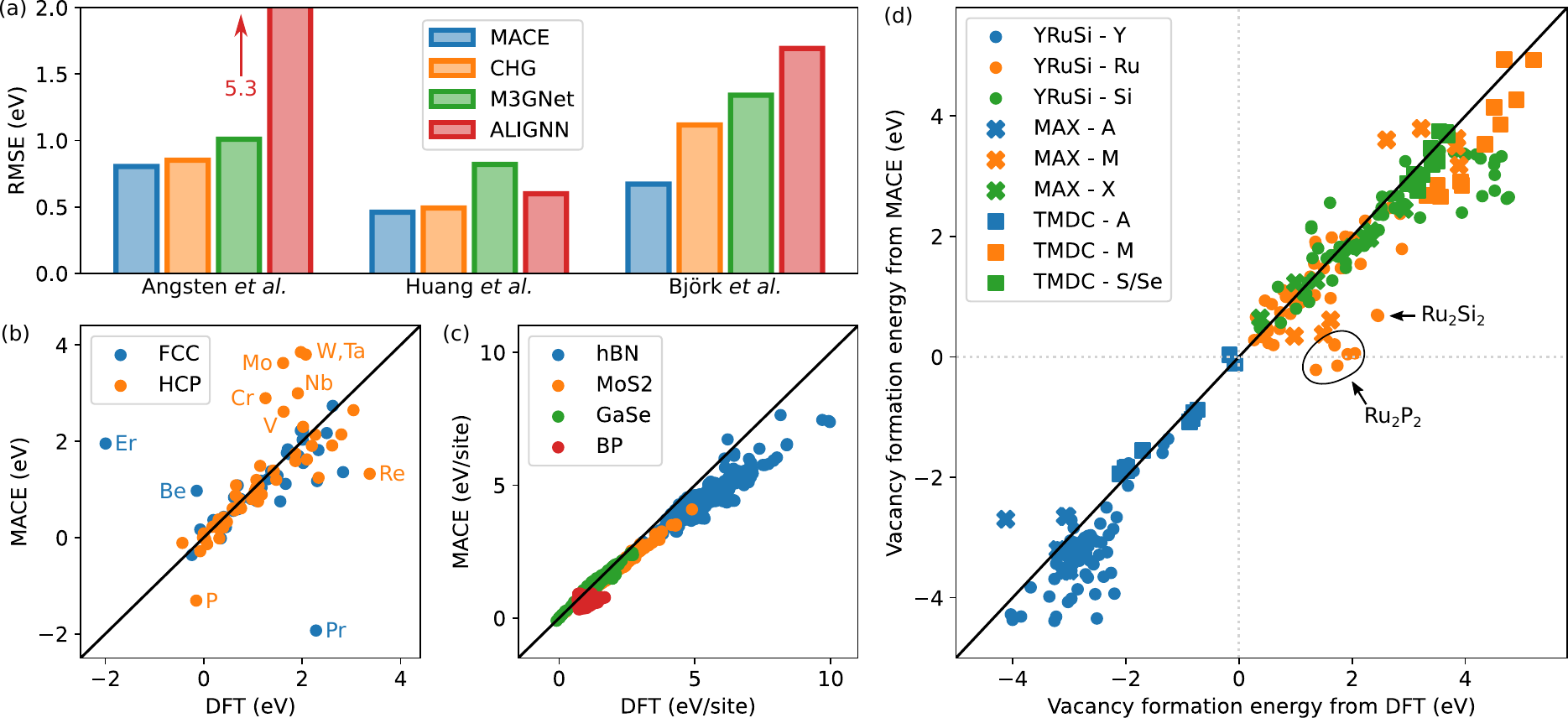}
    \caption{(a) Root-mean-squared error of the UMLIPs tested on data sets of Angsten \etal, Huang \etal, and Björk \etal (Refs. \citenum{Angsten_2014}, \citenum{Huang2023} and \citenum{Bjork2024}, respectively). (b-d) Comparison of the defect formation energy from MACE and DFT for (b) Angsten \etal, (c) Huang \etal, and (d) Björk \etal datasets.}
    \label{fig:Fig1}
\end{figure*}

\subsection{Benchmark of UMLIPs}

In order to assess the accuracy of UMLIPs for the prediction of defect formation energy, we tested four different UMLIPs, namely MACE \cite{Batatia2023foundation}, CHGNet \cite{Deng_2023_chgnet}, M3GNet \cite{Chen2022_m3gnet} and ALIGNN \cite{Choudhary2023_alignn}. These four universal potentials have the advantage of being trained using a similar training set based on the Materials Project database \cite{Jain2013_MP}. All four UMLIPs are tested and compared using three existing defect databases by Angsten \etal\cite{Angsten_2014}, Huang \etal\cite{Huang2023}, and Bj\"{o}rk \etal\cite{Bjork2024}. 
These datasets were chosen because they cover a large variety of materials and because they have been calculated using the same PBE functional \cite{Perdew1996_PBE} as used for training the above-mentioned UMLIPs, and thus we can rule out any error contributions arising from the choice of functional.
The dataset of Angsten \etal contains vacancies of face-centered cubic (FCC) and hexagonal close-packed (HCP) structures for a large part of the periodic table \cite{Angsten_2014}.
The dataset of Huang \etal contains substitutional and vacancy defects in 2D materials, such as h-BN, \ce{MoS2}, GaSe and BP \cite{Huang2023}.
The dataset of Björk \etal contains vacancies for 77 layered transition metal dichalcogenide (TMDC), MAX, and YRuSi-type phases \cite{Bjork2024}.
While the contents of each dataset differ on the structural information given (relaxed, unrelaxed, or no structure) and how the energies are reported (total energy, formation energy and the choice of chemical potentials), our UMLIP calculations are carried out in a way that is consistent with each dataset. Further details concerning how each dataset was processed are given in Sec. I of the SI.

Benchmarking results are presented in Fig.\ \ref{fig:Fig1}. In particular, Fig.\ \ref{fig:Fig1}(a) shows the root mean square error (RMSE) over all three dataset for all four UMLIPs. MACE performs best on all datasets, with RMSE values of 0.80, 0.46 and 0.67 eV for datasets from Angsten \etal, Huang \etal, and Björk \etal, respectively.
Previous benchmarking of UMLIPs for other material properties have also found MACE to extrapolate well to previously unencountered systems \cite{Batatia2022mace,ml_bench_phonon,MLs-bench}.
CHGNet and M3GNet are also found to reasonably predict defect formation energies, although they show larger errors for the layered materials included in the dataset of Björk \etal. On the contrary, ALIGNN seems to perform poorly, which could be explained by the large errors in the chemical potentials (see Fig.\ S9 in the SI). 

More detailed comparison of the MACE and DFT results are given in the correlation plots in Fig.~\ref{fig:Fig1}(b-d). Similar plots for the other UMLIPs can be found in the SI (see Figs. S1, S2 and S3). 
While the agreement for the dataset of Angsten \etal [Fig.\ \ref{fig:Fig1}(b)] is very good for most elements, a few data points are poorly predicted by MACE. For example, HCP phosphorus and FCC erbium and praseodymium show large errors, which are also found when using other UMLIPs (see Fig.\ S1 in the SI). Additionally, transition metals such as molybdenum, tungsten and tantalum also show relatively large errors in the HCP structure, which is not their optimal structure and likely explains the poor performance with MACE.
For the 2D materials, Fig.\ \ref{fig:Fig1}(c), MACE accurately predicts the defect formation energy at low energies. Low formation energy means that the local structural environment is chemically stable and consequently it is likely that similar local environments were already present in the training set for UMLIPs.
While the agreement somewhat worsens at high formation energy, the dependence is still linear, which is important in material screening. 
Largely similar behavior can be observed in Fig.\ \ref{fig:Fig1}(d) for the layered bulk phases included in the database of Björk \etal. In this previous work, the goal was to identify which atoms have a negative vacancy formation energy in acidic solutions and could therefore be etched. In Fig.\ \ref{fig:Fig1}(d), etched atoms are shown in blue and all have $\Delta F_v<0$ according to DFT calculations. It is clear that MACE correctly makes the separation between etched and non-etched atoms, with only a few structures containing ruthenium being wrong. Note that other UMLIPs struggle with the same materials, suggesting that the problem arises from the insufficient training set for these compounds and not from the ML model.

We also carried out benchmarking for datasets including other types of defects, such as substitutions and interstitial atoms, reported in Refs.\ \citenum{Davidsson2023} and \citenum{Rahman2024} (see Fig.\ S4 and S5 in the SI). While the agreement between UMLIPs and DFT calculations remain reasonable for substitutions, it gets clearly worse for interstitial defects. For this kind of defect, DFT formation energy can go up to 20 eV and such highly unstable environments are unlikely to be present in the MP database, resulting in inaccurate predictions from UMLIPs.

Overall, we find that the UMLIPs, and MACE in particular, are successful in predicting vacancy formation energies for a diverse set of materials. 
Although the errors are larger than what is usually considered acceptable when benchmarking UMLIPs, this should not be an issue for the defect screening applications demonstrated below, where correctly predicting the sign of the vacancy formation energy is sufficient. 
Moreover, the accuracy of UMLIPs is comparable to specialized ML models designed to predict formation energies in a limited set of defects/materials: 
0.5--1 eV for oxide defects \cite{Witman2023}, 0.4 eV for neutral oxygen vacancies \cite{Deml15_JPCL}, 0.45--0.7 eV for vacancies in oxide perovskites \cite{Wexler21_JACS}, 0.67 eV for TMDCs \cite{Frey20_ACSNano}, 1 eV for a larger set of defects and materials \cite{Choudhary2023_vac_JARVIS}, 0.27--0.44 eV for vacancies in oxides \cite{Kumagai21_PRM}.
It is important to note that materials with vacancy formation energies close to zero should be treated carefully because of the errors from MACE. In our case, any structure with vacancy formation energy between $-0.75$ and $0.75$ eV is not considered. The value $0.75$ eV is chosen to match the average RMSE of MACE over the benchmarked datasets.

\subsection{High-throughput vacancy calculations}

\begin{figure*}
    \centering
    \includegraphics[width=\linewidth]{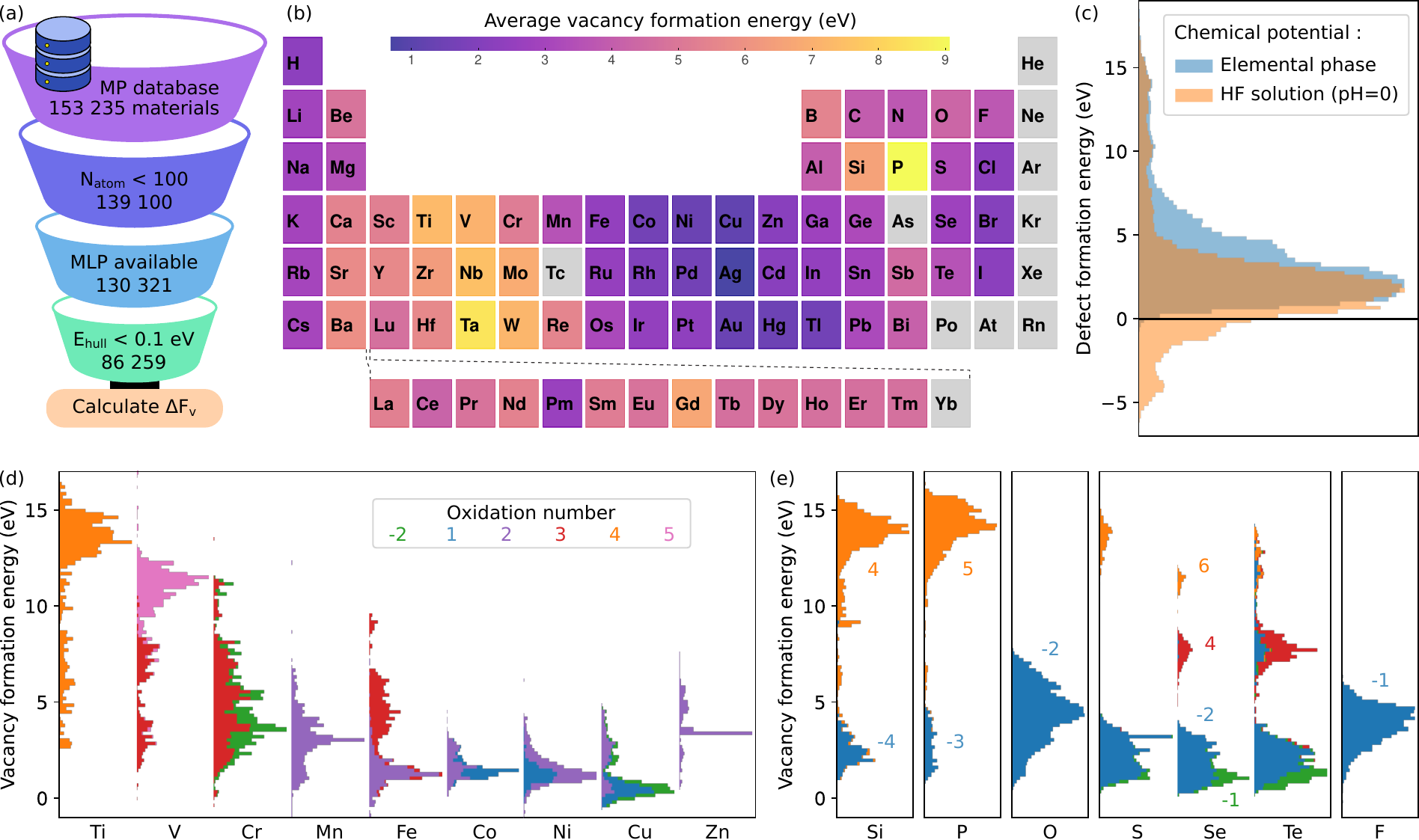}
    \caption{(a) Materials project (MP) database screening procedure. The number of structures in each step is indicated. (b) Average vacancy formation energy for every element in the periodic table. (c) Histograms from all vacancy formation energy calculations with two different choices of chemical potentials. (d,e) Stacked histograms of the vacancy formation energies for the dominant oxidation states of a few selected elements.}
    \label{fig:Fig2}
\end{figure*}

With MACE found to provide reliable vacancy formation energies, we proceeded to calculate vacancy formation for most materials in MP database subject to selected screening conditions. As illustrated in Fig.\ \ref{fig:Fig2}(a), we limit our study to materials with less than 100 atoms in the unit cell, (meta)stable materials corresponding to energy above convex hull less than 0.1 eV/atom, and excluding noble gases, arsenic, technetium, promethium, ytterbium, and elements heavier than Bi (not included in MACE).
This screening still resulted in a total of 86 259 materials.
At this stage, we only calculated the unrelaxed vacancy formation energy due to (i) computational cost, (ii) the ease of analysis, since in some cases relaxation might dramatically alter the structure, and (iii) it provides an upper bound to the formation energy as relaxation can only lower the formation energy.

Fig.\ \ref{fig:Fig2}(b) shows the average vacancy formation energy over the periodic table. 
The lowest formation energies are found for alkali metals, noble metals, and heavier halogens. One generally expects high formation energies with high oxidation state and strong chemical bonds. Here we carry out analysis based mainly on the oxidation state, which is fairly straightforward to extract and a convenient atom-specific scalar quantity. 
Noble metals, by definition, do not like to form compounds and thus low vacancy formation energies are expected. Alkali metals and halogens have oxidation state $\pm 1$, often present as singly charged ions that are relatively easy to extract from the host lattice. Group-11 metals (Cu, Ag, and Au) have a full $d$-shell and a single $s$-electron, and thus they behave largely similarly to alkali metals and with similar average formation energies. On the contrary, high formation energies are found for refractory metals, maximum found for Ta, and also $p$-block elements, with Si and P particularly standing out.

In some cases, the formation energy distribution is multimodal, and thus average formation energy is not sufficiently descriptive. Distributions for few selected elements, and further grouped by their oxidation states, are shown in Fig.~\ref{fig:Fig2}(d-e). 
In SI Table S1 we list the proportion of oxidation states in the database and in SI Table S2 the average formation energies for all elements at all oxidation states.
Oxygen and fluorine both show a very wide distribution, but only existing in the $-2$ and $-1$ oxidation state, respectively.
Si, on the other hand, shows a clear bimodal distribution, dominated by $+4$ and $-4$ oxidation states. The average formation energy for the $+4$ oxidation state is very high at 12.6 eV. A large proportion of these arise from silicates with \ce{SiO4} units, in which removing a silicon atom leaves four undercoordinated O atoms and resulting in high formation energy.
Such high formation energies are not seen in the O distribution, since, while the Si-O bond strength is the same, removing O atom results in only one or two undercoordinated Si atoms and, consequently, the formation energy is also lower by a factor of 2--4. 
The low formation energy peak arises from $-4$ oxidation state, often associated with Si atom surrounded by metals with high coordination numbers (CN). In SI Fig. S6 we show the distributions grouped by CN, which verifies that the high-energy peak is dominated by 4-coordinated Si atoms, while the low-energy peak consists of high-CN Si (largest fraction for CN of 9).

Similar findings apply to P, where bimodal distribution with peaks arising from oxidation states +5 and -3 are observed. The high formation energy peak at 13.7 eV with +5 oxidation state arises mainly from \ce{PO4^3-} units, quantitatively similar to the case of Si, whereas an example of $-3$ oxidation state is phosphine \ce{PH3}. 
SI Fig. S6 shows that the high-energy peak corresponds to 4-coordinated P and 3-coordinated P have very low formation energies. 
Chalcogen elements S, Se, and Te show particularly clear grouping with the oxidation state. These allow us to extract reasonably representative oxidation-state dependent average formation energies. For example, selenium has an average vacancy formation energy of 2.2 eV, 7.5 eV and 11.1 eV in oxidation states -2, +4 and +6, respectively. Detailed results for all elements and all oxidation states can be found in SI Table S2. 

In Fig.~\ref{fig:Fig2}(d) we show the distributions from the $3d$ transition metal series. As mentioned, Cu is found primarily at +1 oxidation state, with a narrow distribution (average 0.6 eV). The lighter elements can adopt a wider range of oxidation states (large oxidation state generally yielding higher formation energies) and thus covering a wider range of formation energies. In few cases, a clear grouping with the oxidation state can be observed, such as \ce{V^{+3}}/\ce{V^{+5}} and \ce{Fe^{+2}}/\ce{Fe^{+3}}. Distributions with CN are given in Fig. S7, but show no obvious grouping. The 4d and 5d series are shown in SI Fig. S8, with largely similar trends.

Overall our results indicate a fairly strong correlation between the oxidation state of the element and its vacancy formation energy. This hints at the possibility of using the oxidation state as a descriptor in carrying out defect screening. Moreover, one can use the data in SI Table S2 as a first approximation for an unknown vacancy formation energy.

\subsection{Ordered vacancy compounds below convex hull}

In the total formation energy distribution, Fig.~\ref{fig:Fig2}(c), one can distinguish a part of the distribution extending to negative energies for 332 materials, 34 of which have vacancy formation energies below $-0.75$ eV.
This would correspond to spontaneous defect formation (when kinetically allowed) until equilibrium is reached by e.g. adopting a stable non-zero vacancy concentration or transforming to another phase.
In other words, these cases hint for a possibility to find new materials that are below the convex hull defined by the materials in the MP database. However, for more quantitative assessment, we need to include three ingredients: exploration of different vacancy concentrations and their ordering (configurations), structural relaxations, and energy comparison to the convex hull.

\begin{figure*}
    \centering
    \includegraphics[width=\linewidth]{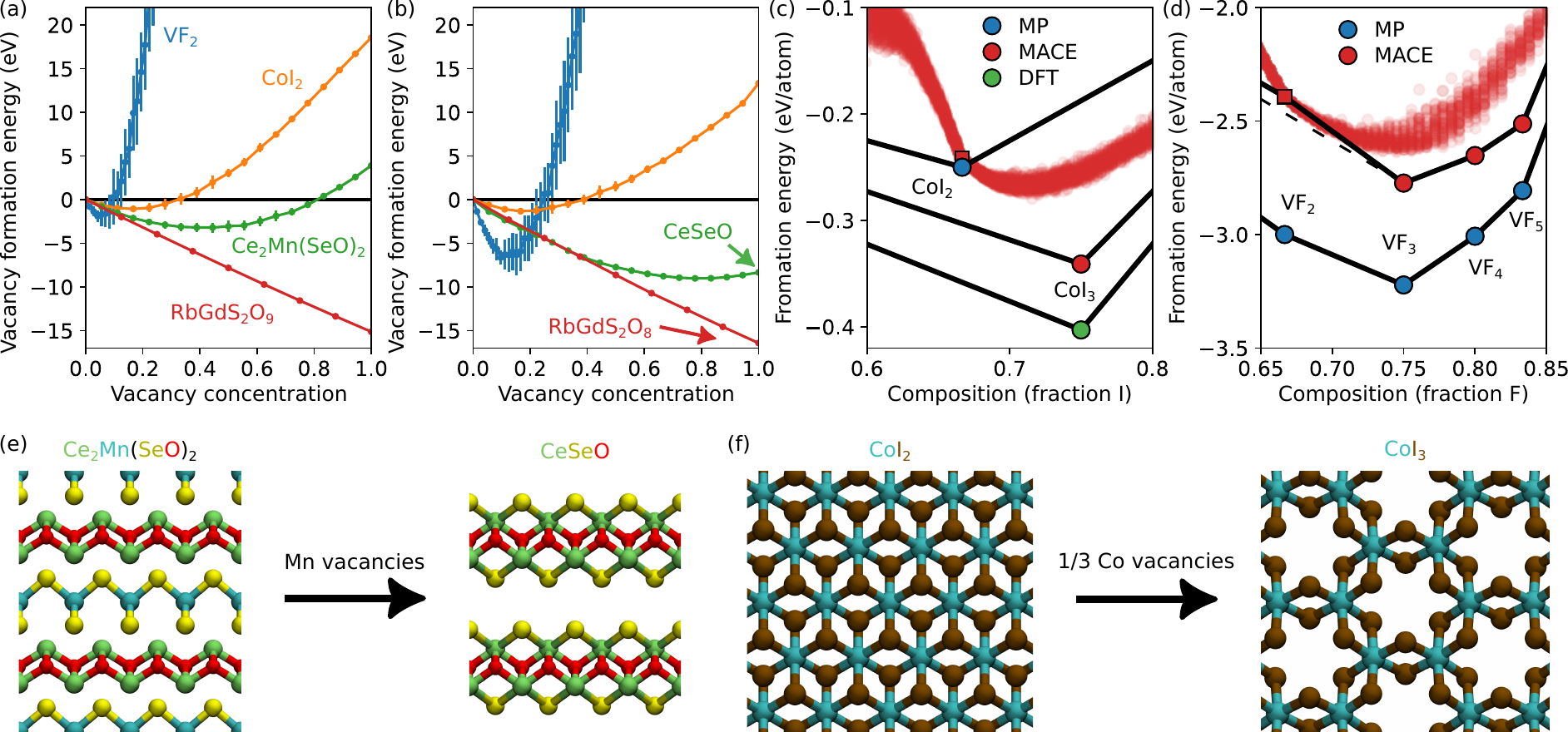}
    \caption{(a) Unrelaxed defect formation energy with respect to the vacancy concentration for a few selected materials. (b) The formation energies after accounting for the structural relaxations. (c-d) Convex hull of \ce{Co-I} and \ce{V-F} systems, respectively. In panels (c) and (d), the red dots show the formation energies of structures containing defects. (e-f) Balls-and-sticks representations of \ce{Ce2Mn(SeO)2} and \ce{CoI2}, respectively. Initial structures from MP are compared with structures with minimal formation energy after removing atoms with negative vacancy formation energies.}
    \label{fig:Fig3}
\end{figure*}

In Fig.~\ref{fig:Fig3}(a), we show the unrelaxed formation energies as a function of defect concentration for four selected materials that were indicated to have negative formation energy and which show qualitatively different behavior.
Except for \ce{VF2}, the deviation of energies around the mean is still small, indicative of weak defect-defect interactions.
\ce{VF2}, \ce{CoI2} and \ce{Ce2Mn(SeO2)2} show minimum formation energy at increasing concentration, with the vacancies created in the V, Co, and Mn sublattices, respectively.
One of the inequivalent O atoms in \ce{RbGdS2O9} is very weakly bound and all are expected to be removed (100\% vacancy concentration in Fig.~\ref{fig:Fig3} referring only to the relevant sublattice), suggesting a possible problem in the experimentally determined structure.

In Fig.\ \ref{fig:Fig3}(b), the same results are shown after relaxation of all relevant systems (including the pristine host). 
One would expect formation energies to decrease and this is indeed found for all cases, although clearly more pronounced in the cases of \ce{VF2} and \ce{Ce2Mn(SeO2)2}. 
Consequently, the vacancy concentration yielding the minimum formation energy increased, from 6 to 14\% in the case of \ce{VF2} and from 44 to nearly 100\% in the case of \ce{Ce2Mn(SeO2)2}.
The structure of \ce{Ce2Mn(SeO2)2} before and after etching all Mn atoms are illustrated in Fig.\ \ref{fig:Fig3}(e). Interestingly, the structure might be intuitively (or by robocrystallographer\cite{Ganose2019}) thought to consist of \ce{Ce2O2} and \ce{MnSe2} layers and thus the possibility to etch Mn atoms would be easily missed.

Having found the lowest energy structures for the defective systems, we can next map their energy with respect to the convex hull.
We emphasize that while the choice of elemental phase references used here is fairly stringent in the sense that it leads to high formation energies, negative defect formation energy does not necessarily imply that the defective phase would be below the convex hull. In fact, the defect formation energy can be directly compared to the slope of the convex hull, and the defective structure falls below the convex hull only if the defect formation energy is lower than the slope of the convex hull.
Co--I and V--F systems both fall into this category.
The relevant part of the phase diagram of Co--I is shown in Fig.\ \ref{fig:Fig3}(c). MP convex hull contains only \ce{CoI2} with the MACE closely reproducing its formation energy.
The calculated systems with Co-vacancies (I fraction above 2/3) fall under the convex hull for a wide range of vacancy concentrations. 
It turns out, however, that the underlying reason for this result is that MP database does not include the stable \ce{CoI3} phase [represented in Fig.\ \ref{fig:Fig3}(f)], although it has been previously predicted computationally by substitution of atoms with chemically similar ones in known lattices \cite{Wang21_npjCM,c2db}.

In the case of V--F, Fig.\ \ref{fig:Fig3}(d), the convex hull in MP consists of \ce{VF2}, \ce{VF3}, \ce{VF4}, \ce{VF5} phases. When recalculated with MACE, the \ce{VF2} phase ends up falling slightly above convex hull. As the \ce{VF3} phase is clearly below \ce{VF2}, the defective system formation energies end up falling very close to the convex hull up to about VF$_{2.5}$, and even slightly below convex hull for some configurations when compared to the \ce{VF2}--\ce{VF3} line. Thus we think it should be possible to synthesize \ce{VF2} with a large range of vacancy concentrations.
A more complete exploration of the possible ordered vacancy compounds would require systematic exploration of all possible supercells and vacancy configurations, and while UMLIPs can accelerate this process, it is beyond the scope of this paper.

For multielement compounds the convex hull is multidimensional and thus difficult to visualize. 
However, the formation energy can still be easily compared with the convex hull of competing phases. \ce{CeSeO} is located on the \ce{CeSe2}--\ce{CeO2} line and is found to be 0.14 eV/atom below the convex hull. Similarly, \ce{RbGdS2O8} has a formation energy 0.23 eV/atom below plane made of \ce{Rb2S2O7}, \ce{Gd2O3} and \ce{SO3}. A structure somewhat similar to \ce{CeSeO} was found by chemical substitution \cite{Wang21_npjCM}, while we are not aware of previous reports for \ce{RbGdS2O8}. 
Thus, these materials could be examples of stable materials below MP convex hull.

\begin{figure*}
    \centering
    \includegraphics[width=\linewidth]{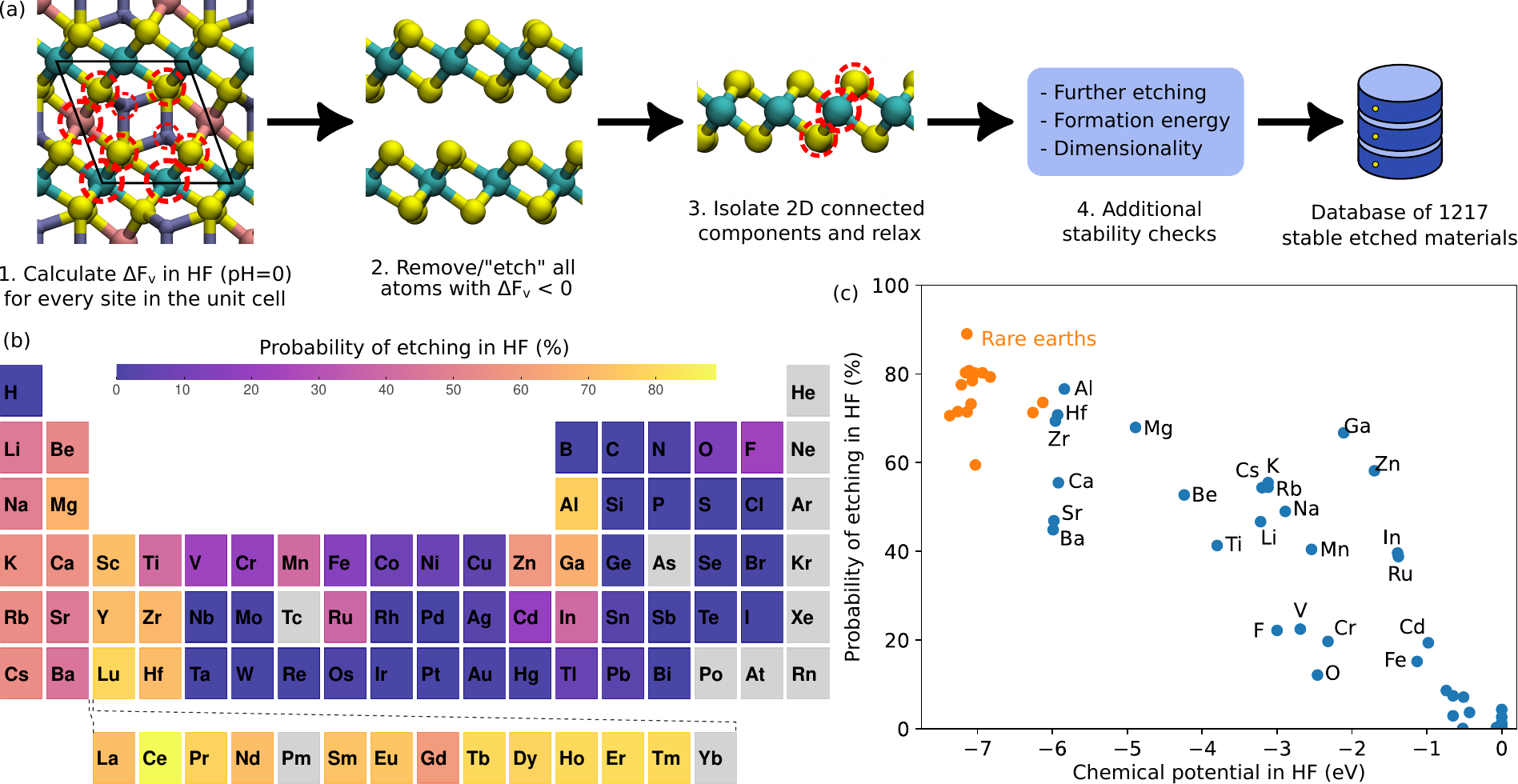}
    \caption{(a) Workflow for the identification of potential 2D layer materials from chemical etching. Dashed red circles indicate the sites in the unit cell for which defect formation energy is calculated.
    (b) Calculated probability of chemical etching in HF for each elements of the periodic table. Excluded elements are shown in gray. (c) Probability of etching with respect to the chemical potential in HF.}
    \label{fig:Fig4}
\end{figure*}

Overall, we have demonstrated how a defect screening can be used to identify known stable materials missing from the database (\ce{CoI3}), possible new stable materials (\ce{CeSeO} and \ce{RbGdS2O8}), as well as systems that are likely to host ordered vacancy compounds (V--F).

\subsection{2D materials from simulated etching}

Another application of vacancy formation energy calculations is the study of chemical etching of bulk phases into low dimensional materials, as previously shown by studying etching of MXenes layers from MAX phases \cite{Bjork2023} and later extended to other layered phases \cite{Bjork2024}. Here, we look for new 2D layers by screening the whole database of vacancy formation energies in a way similar to Björk \etal \cite{Bjork2024}. 

A schematic of the workflow is represented in Fig.\ \ref{fig:Fig4}(a). Starting from the vacancy formation energy database, atoms with negative vacancy formation energy are removed from the atomic structures, with the chemical potentials corresponding to conditions in aqueous HF solution at pH=0.
As can be seen from the histogram in Fig.\ \ref{fig:Fig2}(c), a large portion of materials will have one or more elements etched.
By considering the whole database, one can obtain the etching probability for each element, which is presented in Fig.\ \ref{fig:Fig4}(b). Few groups of high etching probability stand out, mainly consisting of typical cation elements. 
For all these elements, the high etching probability can be explained by a low chemical potential in HF solution, as illustrated in Fig.\ \ref{fig:Fig4}(c). On the contrary, elements with a high chemical potential are rarely etched. Examples would be some transition metals (such as niobium, molybdenum, tantalum and tungsten) and chalcogens other than oxygen. As a result, many of the etched layers presented later contain these elements.

\begin{figure*}
    \centering
    \includegraphics[width=\linewidth]{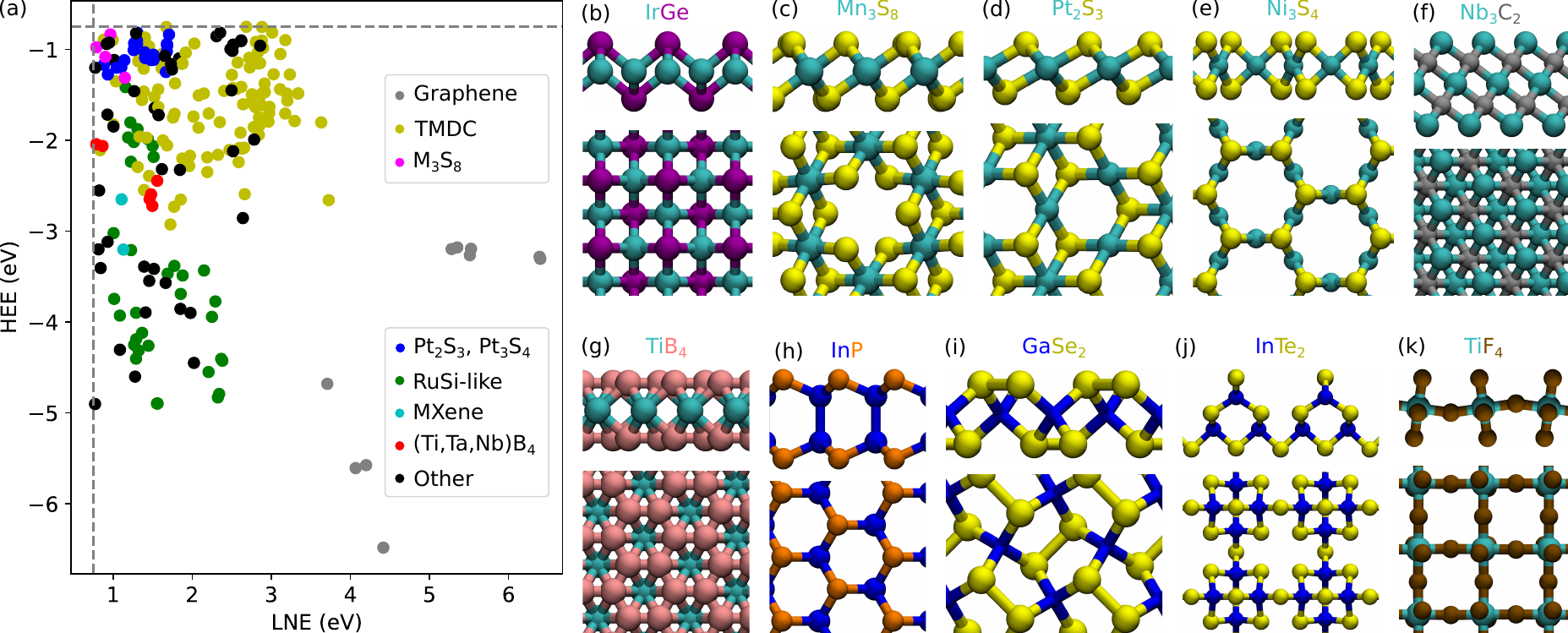}
    \caption{(a) Comparison between the highest etched element (HEE) and lowest non-etched element formation energies (LNE) for binary layers. Note that the plot only shows the region ($\text{HEE}<-0.75,\text{LNE}>0.75)$. Groups of layers with similar atomic structure are indicated by coloring. (b-k) Representation of a few of the best candidate binary layers. Top panels show side view of the layer, while bottom panels show the top view. The color coded formula is given for each structure.}
    \label{fig:Fig5}
\end{figure*}

To find chemically etched 2D materials, layers are first isolated and their atomic structure is relaxed. This procedure resulted in a total of 4370 candidate layers, which are further screened to identify the best candidates. Next, the vacancy formation energies of the layers (denoted $\Delta F_v^{(2)}$, $(2)$ referring to "second round" of etching) are computed. The condition $\Delta F_v^{(2)}>0$ ensures that none of the atoms within the layer is etched. In some cases, relaxing a layer after etching lead to the formation of 1D or 0D clusters. To avoid such cases, the dimensionality of the relaxed layers is calculated again and we only consider those with a 2D score higher than $0.9$. Additionally, layers with positive formation energy are also not considered. After applying these three conditions, we find 1217 candidate layers. In order to identify the best candidates, we define the highest etched element formation energy (HEE) as well as the lowest non-etched element formation energy (LNE) for each layer. Best candidate layers have a low HEE and a high LNE, as well as a high defect formation energy $\Delta F_v^{(2)}$. 
The database is therefore further screened by applying $\text{HEE}<-0.75$, $\text{LNE}>0.75$ and $\Delta F_v^{(2)}>0.75$, resulting in a total of 259 layers labeled as "best candidates".

Fig.\ \ref{fig:Fig5} presents the best candidate unaries and binaries (see Table S3 in the SI for a complete list of all layers). In particular, Fig.\ \ref{fig:Fig5}(a) shows the HEE and LNE of these layers. Many known materials are recovered. 
For example, graphene is found to have the highest LNE as well as the lowest HEE values depending on the initial structure, such as \ce{LiC6} and \ce{EuC6}, where lithium and europium are respectively etched. Note that graphene is also the material with the highest value for $\Delta F_v^{(2)}$.
Among layers with low HEE, there are many with an atomic structure similar to \ce{RuSi} recently found by Björk \etal \cite{Bjork2024}.
Fig.\ \ref{fig:Fig5}(a) also shows two MXene structures, namely \ce{Nb3C2} and \ce{Ta3C2} [see Fig.\ \ref{fig:Fig5}(f)]. Although these two MXenes have not been synthesized, there has been report of successful synthesis of other similar layers based on Nb and Ta \cite{Anasori2022}.

A large fraction of the best candidates are TMDCs, ranging from well known \ce{MoS2}, \ce{NbS2} or \ce{TaSe2} \cite{Manzeli2017} to more exotic \ce{RhSe2} or \ce{RhTe2}. Interestingly, TMDCs can be obtained by etching a large variety of elements. For example, \ce{NbS2} can be found by etching alkali metals such as Li and Cs, early transition metals such as Ti and V or post-transition metals such as In.
In addition to simple TMDCs, we also find TMDC layers including defects, either metal or chalcogen vacancies. An example of metal vacancy would be \ce{Mn3S8} layers, which is represented in Fig.\ \ref{fig:Fig5}(c). Note that these layers are found by etching two different elements. For chalcogen vacancies, we find two different cases, namely \ce{Pt2S3} and \ce{Pt2Se3}. An example of the atomic structure can be seen in Fig.\ \ref{fig:Fig5}(d). Similar ordered vacancies have already been experimentally realized in \ce{PtTe2} \cite{Xu2024}.
The final example of binary materials containing transition metals and chalcogens is \ce{Ni3S4}, \ce{Pd3Se4} or \ce{Pt3S4}. All these layers have a Kagome structure similar to the one shown in Fig.\ \ref{fig:Fig5}(e), and are obtained by etching alkali metals from a parent phase resembling \ce{Cs2Ni3S4}. Previous experimental study have already highlighted the possibility to etch Cs atoms from \ce{Cs2Ni3S4}, although they could not reach \ce{Ni3S4} monolayers \cite{Villalpando2024}.

The screening also revealed semiconducting layers, such as \ce{InP} [Fig.\ \ref{fig:Fig5}(h)], which has already been experimentally synthesized using chemical etching \cite{Bae2024} and has the same atomic structure as the well known \ce{GaSe} \cite{Hu2012,Cao2015}. 
Although \ce{GaSe} is not included in our results, a different form of gallium selenide is, namely \ce{GaSe2}. This layer takes a pentagonal atomic structure similar to \ce{PdSe2} \cite{Oyedele2017,Chen2020,Sierra2025} and is represented in Fig.\ \ref{fig:Fig5}(i).
Another new semiconducting layer is \ce{InTe2} [see Fig.\ \ref{fig:Fig5}(j)], etched from \ce{CsInTe2}. Although such parent phases are known to be present in alkali-metal treated CIGS solar cells, there are no reports on isolating them as monolayers. Note that we also find other compositions with the same structure, such as \ce{InSe2} or \ce{GaTe2}, as well as \ce{InTe2} in the pentagonal form [similar to \ce{GaSe2} in Fig.\ \ref{fig:Fig5}(i)], although these examples do not meet the criteria to be among the "best candidates".

Among the other lesser known materials, we first highlight \ce{TiB4}, \ce{NbB4} and \ce{TaB4}. These layers are composed of two hexagonal boron layers, with metal atoms located in-between at the center of the hexagons [see Fig.\ \ref{fig:Fig5}(g)]. The parent phases are similar to bulk \ce{TiB2} \cite{Basu2006}, but with alternating layers of Ti and either Zr or Hf, and the latter are etched to lead to \ce{TiB4}. Thicker layers, such as \ce{Nb2B6} or \ce{Ta3B8}, are also predicted. Note that these layers resemble MBenes,
which are absent from our results due to terminations not being included. Finally, bulk materials such as \ce{CsTiF4} are found to be etchable to form \ce{TiF4}. The atomic structure of this layer consist of corner sharing \ce{TiF6} octahedra [Fig.\ \ref{fig:Fig5}(k)]. Although very similar to other titanium fluoride structures \cite{Davidovich2015}, the stability of \ce{TiF6} in 2D form remains unknown.

\begin{figure*}
    \centering
    \includegraphics[width=\linewidth]{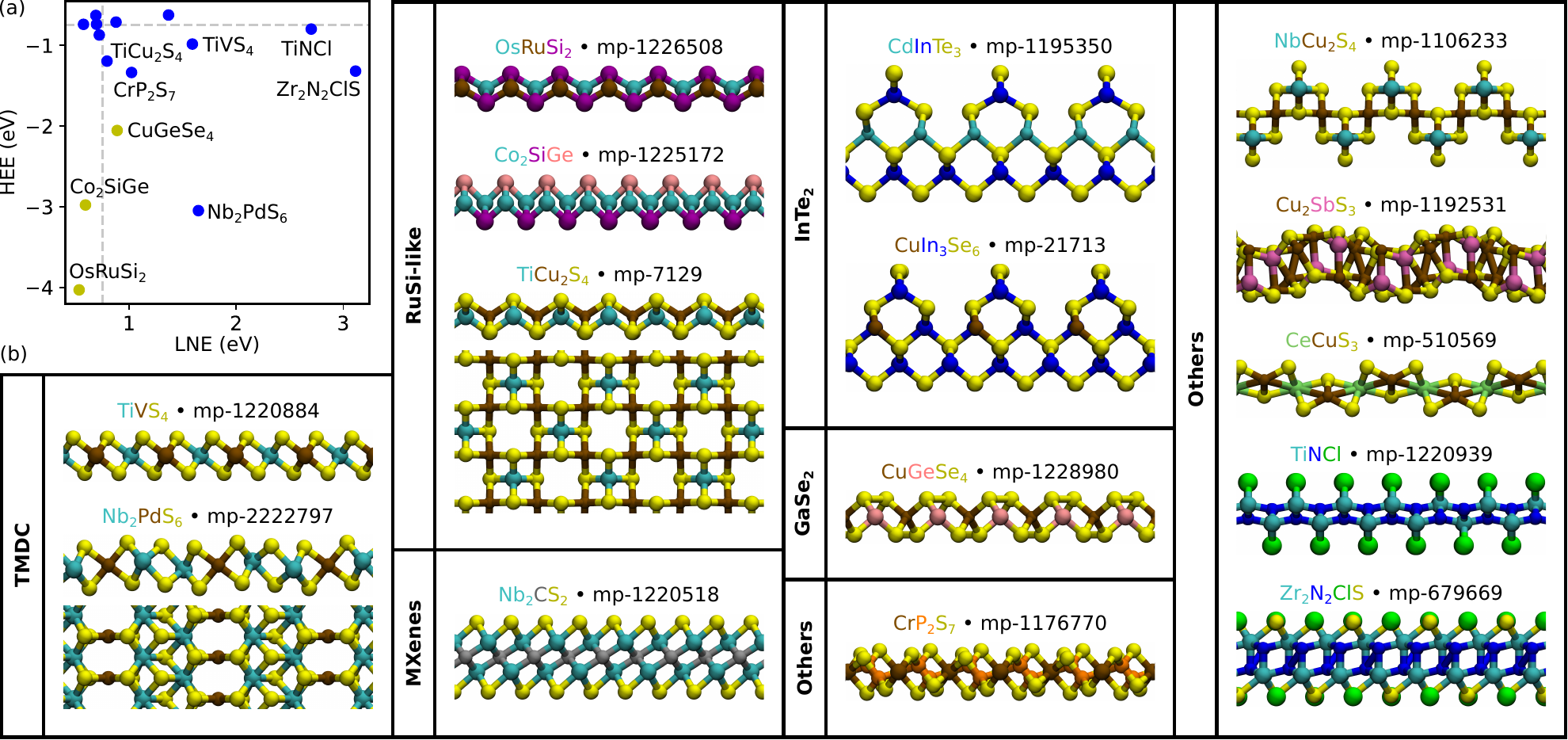}
    \caption{(a) Comparison between LNE and HEE for ternary materials. Grey dashed lines show the limits for LNE$=0.75$ and HEE$=-0.75$. Blue dots represent layers with $\Delta F_v^{(2)}>0.75$, while yellow ones have $0.5<\Delta F_v^{(2)}<0.75$. (b) Representation of a few of the best candidate ternary layers. The structures are grouped depending on their resemblance with binary layers (see Fig.\ \ref{fig:Fig5} for the structure of binary layers). The color-coded formula and the Materials Project ID (MPID) is given for each structure.}
    \label{fig:Fig6}
\end{figure*}

While previous works by Björk \etal only focused on binary layers etched from ternary parent phases, there was no need to limit the number of elements in our study due to the high performance of UMLIPs. Hence our calculations also predicted ternary (and quaternary) layers after etching, which are presented in Fig.\ \ref{fig:Fig6}. A complete list of ternary layers is available in Table S4 of the SI. In particular, Fig.\ \ref{fig:Fig6}(a) compares the LNE and HEE of the best candidates. Note that here, the threshold previously set at 0.75 is lowered to 0.5, resulting in a total of 77 layers (see table S4 in the SI). The atomic structures of a few of the best candidate ternary layers are presented in Fig.\ \ref{fig:Fig6}(b). Some of these structures are alloyed variants of the known binary structures, such as TMDC \ce{TiVS4}. In \ce{Nb2PdS6}, Nb has 7 neighboring S atoms while Pd only has 4, which leads to an atomic structure different from the usual TMDCs. We also find alloys of RuSi-like layers, such as \ce{OsRuSi2} which has alternating elements on the Ru layer, or \ce{Co2SiGe} which has a Janus structure. In addition to alloying, some showed ordered vacancies, such as \ce{TiCu2S4}, with alternating Ti and Cu atoms and Ti vacancies on half of the sites. Two other similar structures were found, namely \ce{TiCu2Se4} and \ce{TiAg2S4}. 

Regarding MXenes, ternary layers \ce{Nb2CS2} and \ce{Ta2CS2} are included among the best candidates. These layers have the structure of MXenes with sulfur atoms as terminations and have already been synthesized, showing promising applications as superconductors \cite{Kamysbayev2020,Majed2022}. Note that contrary to typical MXenes where aluminum atoms are etched, our calculation suggest that \ce{Nb2CS2} and \ce{Ta2CS2} can be obtained by etching 3d transition-metals from V to Cu.

Ternaries similar to \ce{InTe2} and \ce{GaSe2} are also found. For the former, it has the same atomic structure as represented in Fig.\ \ref{fig:Fig5}(j) but with an additional layer of \ce{InTe} and alloying on the middle In sublattice. This leads to two different structures with different concentration of alloys, namely \ce{CdInTe3} and \ce{CuIn3Se6}, which are both shown in Fig.\ \ref{fig:Fig6}(b). Regarding \ce{GaSe2}, two similar pentagonal structures with alternating elements on the gallium site were found, namely \ce{CuGeSe4} and \ce{AgSnSe4}. 

In addition to ternaries resembling binary layers, our results also highlight ternary layers with completely new atomic structures, including \ce{CrP2S7}, \ce{Cu2SbS3} or \ce{NbCu2S4}. Note that for the latter, the same atomic structure is found for a wide variety of compositions, such as \ce{TaAg2Se4} or \ce{VCu2S4} (see Table S4 in the SI for a complete list). 
While rare earth metals are found to have a very high probability of being etch in HF, \ce{CeCuS3} is found to be among the candidates for chemical exfoliation. Other layers sharing the same atomic structure are also among candidates, although they have lower LNE and $\Delta F_v^{(2)}$ values, for example \ce{CoTmS3}. In addition to rare eath metals, the same structure is also found in layers containing two transition metals, such as \ce{CuTiS3} and \ce{CuZrS3}.

A majority of materials presented so far include chalcogens at the surface. This can be explained by their low probability of etching [see Fig.\ \ref{fig:Fig4}(b)], which would tend to stabilize layers. Among ternary layers, we find two candidates with chlorine terminated surfaces, namely \ce{TiNCl} and \ce{ZrNCl}. These metal nitride halides are known to take two possible atomic structures, one cubic and one hexagonal \cite{Osanloo2022}, both represented in Fig.\ \ref{fig:Fig6}(b). Although nanosheets of \ce{ZrNCl} have been synthesized by mechanical exfoliation \cite{Feng2015}, there is no report of chemical etching for this material. Note that for \ce{ZrNCl}, our results also contain layers where half of the chlorine atoms are replaced by sulfur, leading to \ce{Zr2N2ClS}. In particular, we find two structures with this formula, one with alternating Cl-S atoms and one with a Janus type.

\section{Discussion}

In conclusion, calculations of vacancy formation energies contain useful information about materials properties, yet they are computationally demanding and usually restricted to a small amount of structures. We showed how UMLIPs can be used to drastically reduce the computational cost while maintaining a good accuracy. After carefully benchmarking four different UMLIPs, MACE was found to yield the most accurate predictions. It was then used to compute vacancy formation energies of 86 259 materials, giving access to large statistics for most elements in the periodic table. As an application, vacancy formation energies were used to explore new ordered vacancy compounds near the convex hull and the synthesis of new two-dimensional layers using chemical etching. 
Most the ordered vacancy compounds below or near convex hull identified in this work were indicative of known compounds missing from the databases. Nevertheless our work brings forth the importance of defects in the exploration of new materials, a dimension that has been previously largely ignored. 
That said, given the small energy differences near the convex hull, a more quantitative identification would benefit from more accurate UMLIPs, especially when defects other than vacancies are considered. We hope that this is achievable in the future, given the fast progress in the development of UMLIPs. At the same time, we recommend including defective materials in the training process to further accelerate their use in defect screening. 

Making use of the high efficiency of UMLIPs, this work predicted 1217 two-dimensional layers obtainable through chemical etching. These candidate layers include well known materials, as well as new exciting ones. 
Importantly, our approach could be readily applied to materials with any number of elements, any number of atoms in the unit cell, any crystal symmetry, and to cases where several elements may be etched. It is important to note that the current work did not investigate the stability of these layers, nor tested their experimental synthesis. Additional work is therefore required in order to validate some of the current predictions, as well as identify the properties of these new layers and their potential future applications. 
While surface terminations play an important role in the stability of some 2D layers, they have not been considered in this work. Including terminations would lower the formation energy and the vacancy formation energy of 2D layers. As a result, more layers would pass the screening depicted in Fig.\ \ref{fig:Fig4}(a). For example, \ce{Ti3C2} is a well known experimentally synthesized MXene layer, but not considered in this work because of the negative vacancy formation energy 
of titanium atoms. Testing various elements at different surface sites for a large number of structures is not straightforward and would prove challenging, hence the decision not to include terminations in the current work. The database containing etched layers could however be a good starting point for future studies on surface terminations of 2D layers. 
Additionally, general trends can also be observed using the large statistics available from MP materials. For example, Fig.\ \ref{fig:Fig4}(b) shows which elements are most likely to be etched, and which ones are more stable in HF. These results can be further used for the design of new low-dimensional materials and their parent bulk phase outside of the MP database.

\section{Methods}

\subsection*{Universal machine-learning interatomic potentials}

We used four different UMLIPs, namely MACE \cite{Batatia2023foundation} (MACE-MP-0, small model size, without dispersion), CHGNet \cite{Deng_2023_chgnet} (default model from the chgnet python package), M3GNet \cite{Chen2022_m3gnet} (M3GNet-MP-2021.2.8-PES model) and ALIGNN \cite{Choudhary2023_alignn} (mp\_traj model). Examples of python code for each UMLPs are given in the SI Section V. All the considered UMLIPs are available within the atomic simulation environment (ASE) \cite{ase-paper}, which is used to run all the calculations.

\subsection*{Formation energies}

The formation energy of the host material $F$ can be defined as
\begin{equation}
    F = E - \sum_{i}\mu_i,
    \label{equ:formation_energy}
\end{equation}
where $E$ is the total energy and $\mu_i$ is the chemical potential of atom $i$. In the context of phase diagrams (showing convex hull), the chemical potentials are taken from the most stable elemental phase. 

The defect formation energy $\Delta F_d$ can be defined as the difference between the formation energy of the defective system $F_d$ and the one from the pristine cell $F_0$, which reads
\begin{equation}
    \Delta F_d = F_d - F_0.
    \label{equ:defect_formation_energy}
\end{equation}
While equation \ref{equ:defect_formation_energy} works for any defective structure, including substitutional defects in Huang \etal dataset, our work mainly focuses on vacancies. In this case, the vacancy formation energy $\Delta F_v$ can be written directly using total energies
\begin{equation}
    \Delta F_v = E_d + \mu_A - E_0,
    \label{equ:vacancy_formation_energy}
\end{equation}
where $E_d$ and $E_0$ denote the total energy of the defective and pristine cells, respectively. The accuracy of UMLIPs to predict $\Delta F_v$ will depend on how well they can reproduce the three energy terms. 
Correlation plot of the total energies for the elemental references between DFT results in MP and those predicted using UMLIPs are shown in SI Fig. S9.

The chemical potentials must be carefully chosen to reflect the environment the material is in, i.e., the reservoir with which atoms are exchanged.
The elemental phase reference is often adopted in the literature due to convenience and due to the consistency with the phase diagrams.
When considering etching if HF, we use the values previously used in Ref.\ \citenum{Bjork2024}. 
The values are taken from experiments, but shown to yield good predictions for the probability of etching of MAX phases \cite{Bjork2023}. 
We supplement their list with the values for lanthanides when available, taking the experimental values from the NBS tables \cite{NBS_table_1982} and adding the same corrections as in Ref.\ \citenum{Bjork2024}.
The chemical potential values are listed in SI Table S5 and S6.

\subsection*{High-throughput vacancy calculations}

We extracted conventional unit cells of materials from MP (v2023.11.1) \cite{mp} 
using MP API \cite{mpapi} and Pymatgen \cite{pymatgen}. Supercells were constructed by repeating the unit cell from MP until the lattice constants were larger than 10 {\AA}. The total energy $E_0$ of the pristine supercell is then calculated using MACE, as well as the total energy of the supercell with a vacancy $E_d$ for every site in the unit cell. The vacancy formation energy is then obtained from equation \ref{equ:vacancy_formation_energy} and saved in a database available at \cite{berger_2025_15025795}. Note that while the analysis presented in this work focuses on materials close to the convex hull ($E_\text{hull}<0.1$ eV), the database also contains vacancy formation energies for materials above the convex hull, leading to a total of 130 321 entries. The database also contains the vacancy formation energy using chemical potentials in the elemental phase as well as in HF solution.

The oxidation states were extracted from MP \cite{mp,mpapi}. The coordination numbers were determined using CrystalNN algorithm in Pymatgen \cite{crystalnn}.

\subsection*{Defective materials below convex hull}

For \ce{VF2}, \ce{CoI2}, \ce{Ce2Mn(SeO)2} and \ce{RbGdS2O9}, high density of defects were also investigated. Supercell were created following the same method as in high-throughput vacancy calculations, except for \ce{CoI2}, where larger 6x6x3 supercells were used. Vacancies are created by removing atoms with negative $\Delta F_v$ at random. For each concentration, 50 supercells containing vacancies are created and the formation energy is computed before and after relaxation. 

Convex hulls from MP were obtained using the MP API \cite{mpapi}. It is important to note that anion corrections are applied to formation energies in MP \cite{Wang2021}. Same corrections were also applied here when calculating convex hull using MACE or DFT calculations.

\subsection*{2D materials from simulated etching}

From the vacancy formation energy database, 2D layers are obtained by removing atoms with negative vacancy formation energy. Resulting layers with 2D dimensionality higher than 0.5 are then isolated and relaxed. The assessment of the dimensionality and isolation of the layers are done using methods already implemented in the ASE \cite{Larsen2019,ase-paper}. Relaxation is performed using the BFGS algorithm with forces and energies calculated with MACE. All 8017 resulting layers are saved in a database available at \cite{berger_2025_15025795}. Note that this database also contains layers etched from parent materials above the convex hull ($E_\text{hull}<0.1$ eV), leading to a higher number of structures than discussed in the text.


\section*{Aknowledgment}
We are grateful to the Research Council of Finland for support under Academy Project funding No. 357483.
The authors thank CSC–IT Center for Science Ltd. for generous grants of computer time.

\section*{Conflict of Interest}

The authors declare no conflict of interest.

\section*{Data Availability Statement}

The vacancy formation energy from high-throughput calculations are saved in a database available at \cite{berger_2025_15025795}. All 8017 layers from the simulated etching are saved in a database available at \cite{berger_2025_15025795}. 
Other data and scripts used in this study are available from the corresponding author upon reasonable request.

\bibliography{Vac_MACE}

\end{document}


\author{Ethan Berger}
\affiliation{Microelectronics Research Unit, Faculty of Information Technology and Electrical Engineering, University of Oulu, P.O. Box 4500, Oulu, FIN-90014, Finland}
\affiliation{Department of Physics, Chalmers University of Technology, SE-41296 Gothenburg, Sweden}
\author{Mohammad Bagheri}
\affiliation{Nanoscience Center, Department of Physics, University of Jyv\"askyl\"a, Finland}
\author{Hannu-Pekka Komsa}
\affiliation{Microelectronics Research Unit, Faculty of Information Technology and Electrical Engineering, University of Oulu, P.O. Box 4500, Oulu, FIN-90014, Finland}
\email{hannu-pekka.komsa@oulu.fi}

\title{Supporting Information: \\ Screening of material defects using universal machine-learning interatomic potentials}
\maketitle

\section{Benchmark of UMLPs} 

\subsection{Angsten \etal}

\begin{figure*}[h]
    \centering
    \includegraphics[width=0.8\linewidth]{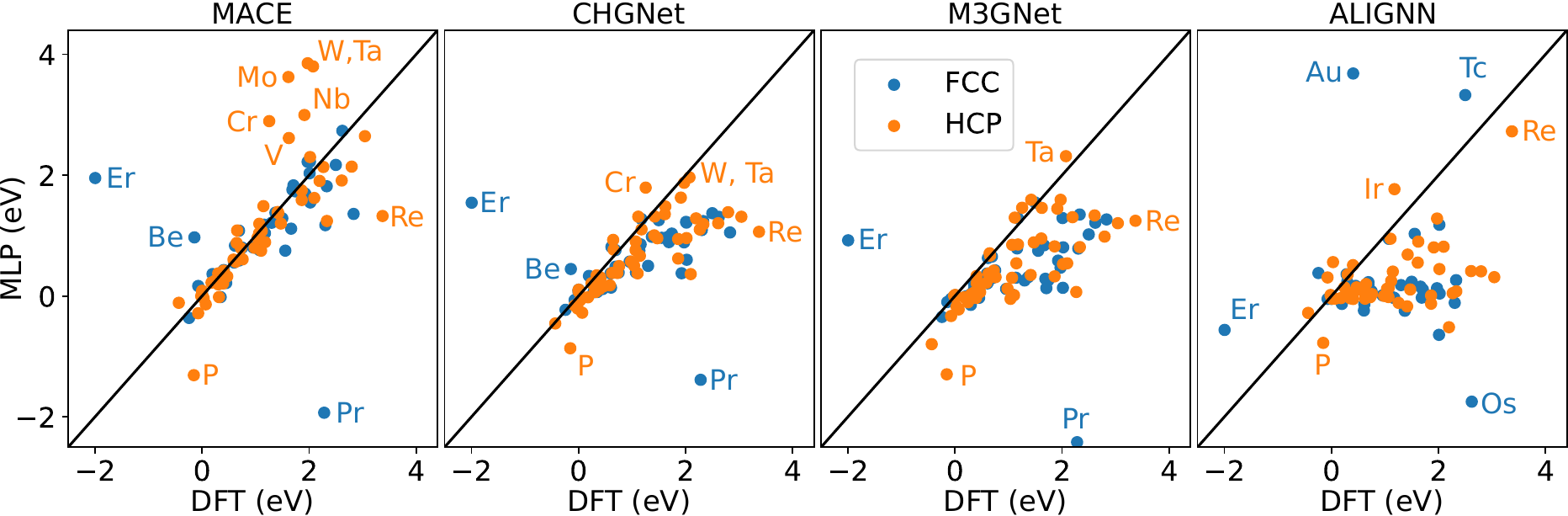}
    \caption{Comparison of the defect formation energy from DFT calculations with (a) MACE, (b) CHGNet, (c) M3GNet and (d) ALIGNN. Reference DFT values are taken from Angsten \etal \cite{Angsten_2014}.}
    \label{fig:Benchmark_Angsten}
\end{figure*}

This dataset contains FCC and HCC structures for most of the periodic table. Outputs of the DFT calculations including defects are directly available at Ref.\ \citenum{Angsten_2014_repository}, making the comparison straightforward. Reference vacancy formations energies are directly obtained using the energy in the output file, while the defective structures are used to compute the vacancy formation energy with UMLPs. Results for all four UMLPs are shown in Fig.\ \ref{fig:Benchmark_Angsten}.

\subsection{Huang \etal}

\begin{figure*}[h]
    \centering
    \includegraphics[width=0.8\linewidth]{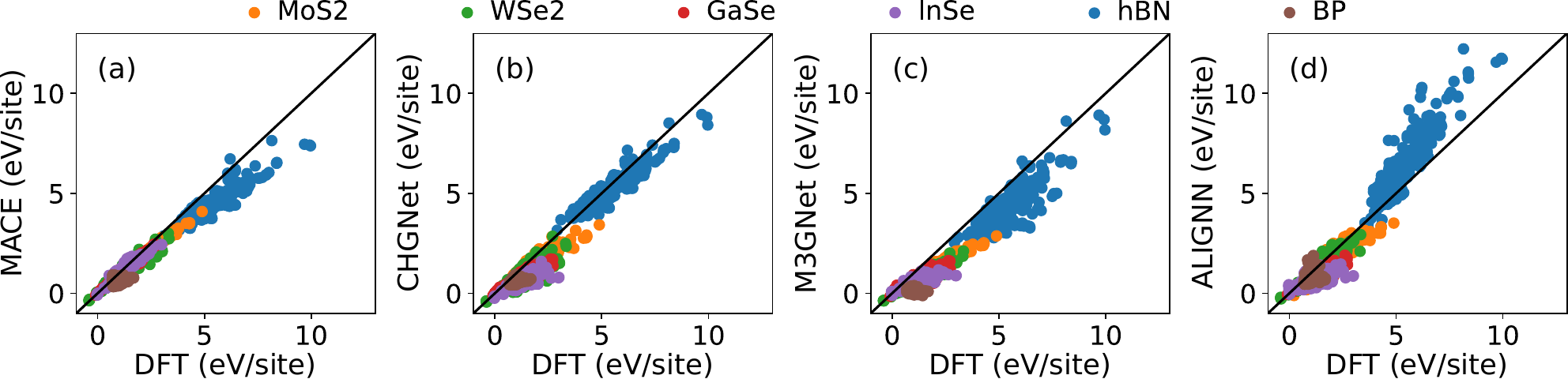}
    \caption{Comparison of the defect formation energy from DFT calculations with (a) MACE, (b) CHGNet, (c) M3GNet and (d) ALIGNN. Reference DFT values are taken from Huang \etal \cite{Huang2023}.}
    \label{fig:Benchmark_Huang}
\end{figure*}

Dataset from Ref.\ \citenum{Huang2023} contains defect formation energies for 6 different 2D layers with high density of defects (vacancies and substitutions). Note however that only the initial unrelaxed defective structures are shared. These are therefore first relaxed using MACE, and the defect formation energy of the resulting relaxed structures is then computed using all four UMLPs. Results are shown in Fig.\ \ref{fig:Benchmark_Huang}. Note that the good agreement between MACE and DFT proves that structures were correctly relaxed.

\newpage

\subsection{Björk \etal}

\begin{figure*}[h]
    \centering
    \includegraphics[width=0.8\linewidth]{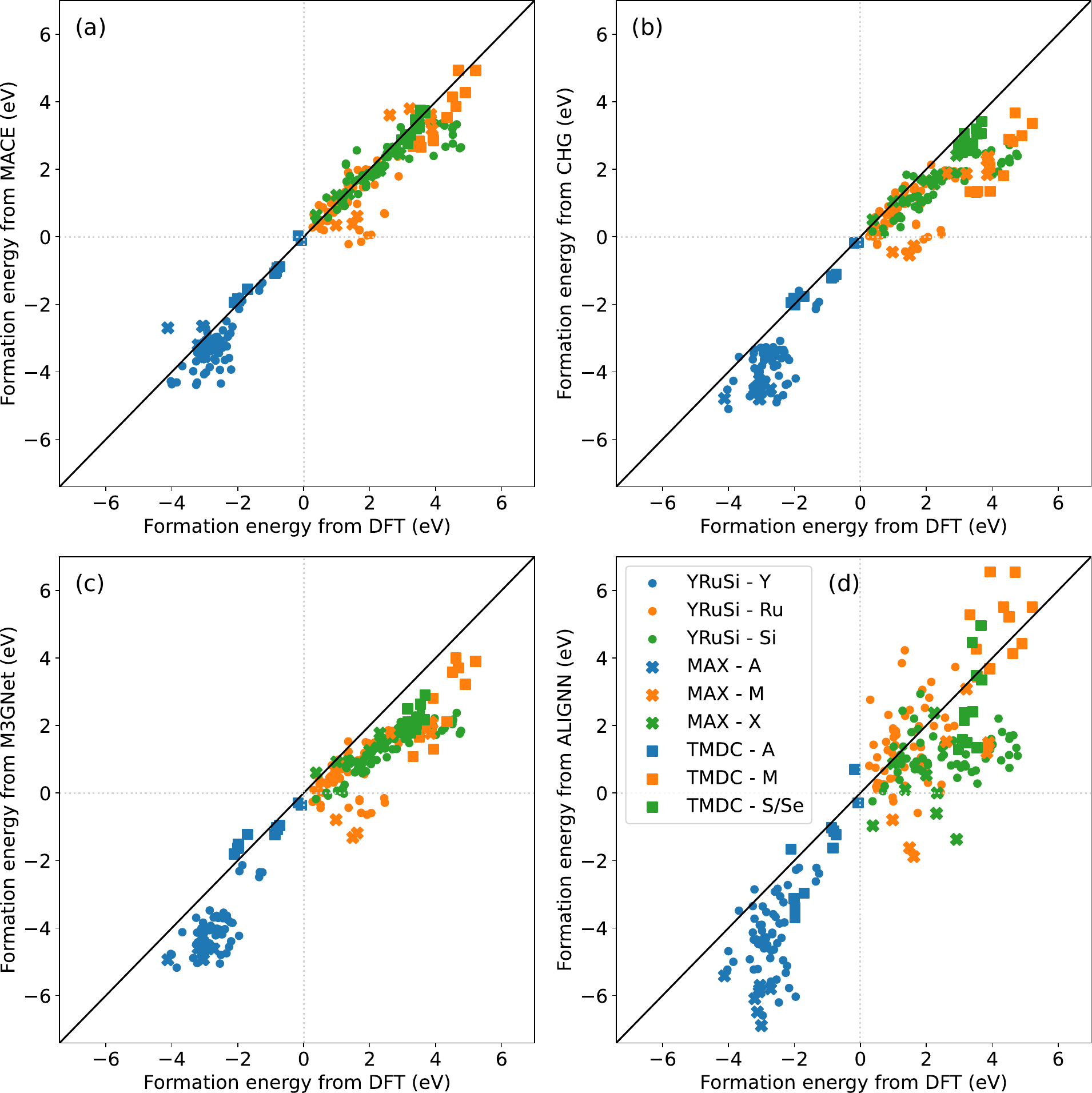}
    \caption{Comparison of the defect formation energy from DFT calculations with (a) MACE, (b) CHGNet, (c) M3GNet and (d) ALIGNN. Reference DFT values are taken from Bj\"{o}rk \etal \cite{Bjork2024}.}
    \label{fig:Benchmark_Bjork}
\end{figure*}

In Ref.\ \citenum{Bjork2024}, Björk \etal screened the Materials Project database in a similar way to what is presented in the main text. As a result, they obtain a database with defect formation energies for every sites of many materials, which can be grouped into three main families : TMDCs, MXenes and RuSi-like. For TMDCs, we used either \ce{LiNbS2} or \ce{InNbS2} as a starting layer, depending if the A sites were alkali metals or post-transision metals. Other structures were then created by substitution of the elements and relaxation using MACE. Similarly, for MXenes and RuSi-like, we started with \ce{Ti_{n+1}AlC_n} and \ce{YRu2Si2} and performed the same substitution and relaxation to obtain the rest of the structures. Defect formation energies were then calculated using the final relaxed structures and all four UMLPs. A comparison between DFT results and UMLPs predictions is shown in Fig.\ \ref{fig:Benchmark_Bjork}. Here again, note that the good agreement between MACE and DFT proves that structures were created correctly.

\newpage

\subsection{Other datasets}

\begin{figure*}[h]
    \centering
    \includegraphics[width=0.8\linewidth]{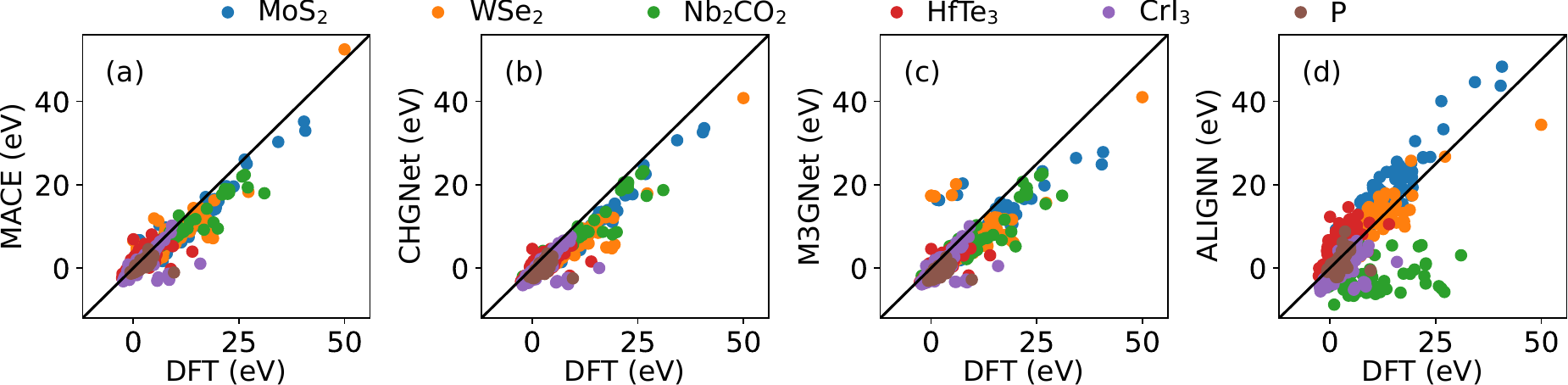}
    \caption{Comparison of the defect formation energy from DFT calculations with (a) MACE, (b) CHGNet, (c) M3GNet and (d) ALIGNN. Reference DFT values are taken from Ref.\ \citenum{Davidsson2023_db}.}
    \label{fig:Benchmark_2DMD}
\end{figure*}

\begin{figure*}[h]
    \centering
    \includegraphics[width=0.8\linewidth]{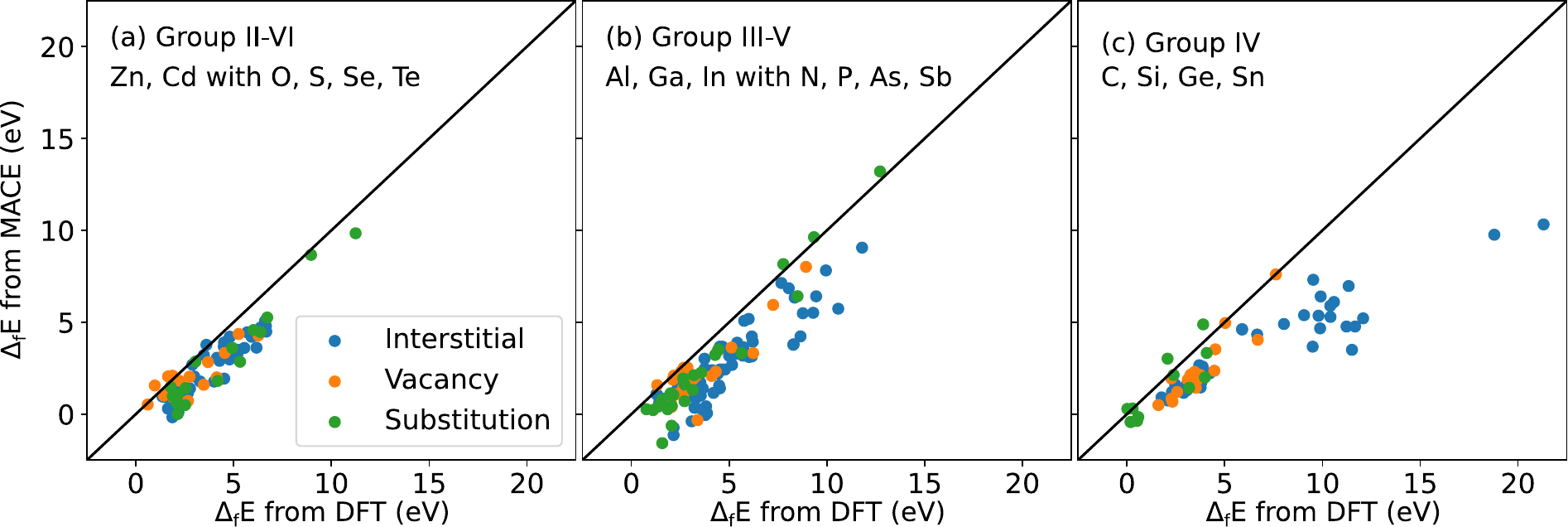}
    \caption{Comparison of the defect formation energy from DFT calculations with MACE in various semiconductors. Different types of defects are separated by colors. Reference DFT values are taken from Ref.\ \citenum{Rahman2024}.}
    \label{fig:Benchmark_Choudary}
\end{figure*}

In Fig.\ \ref{fig:Benchmark_2DMD} we show the results from the defect database of Davidsson \etal \cite{Davidsson2023_db}. It contains interstitial and adatom impurities on 2D materials. 
Finally, in Fig.\ \ref{fig:Benchmark_Choudary} we show the results from the defect database of Choudhary \etal \cite{Rahman2024},
which covers vacancy, substitutional, and interstitial defects in few common semiconductors and insulators.
We think that the downshift of the UMLP values might arise from the adopted chemical potentials, for which we could not find proper documentation. 

\newpage

\section{Oxidation states and coordination number}

Fig.\ 2(d) and (e) of the main text show the distribution of vacancy formation energy for different oxidation states of various elements. The oxidation states were obtained using pymatgen and their distribution is presented in Table \ref{tab:oxi}. For some elements, there are many possible oxidation states. In Fig.\ 2(d-e), we only considered oxidation states with ratio higher than 20\%, which are shown in bold in the following tables.

\begin{longtable}{lccccccccccccc}
\hline
 \hspace{20pt}  & \hspace{7pt}  -4 \hspace{7pt} & \hspace{7pt} -3 \hspace{7pt} & \hspace{7pt} -2 \hspace{7pt} & \hspace{7pt} -1 \hspace{7pt} & \hspace{7pt} 0 \hspace{7pt} & \hspace{7pt} 1 \hspace{7pt} & \hspace{7pt} 2 \hspace{7pt} & \hspace{7pt} 3 \hspace{7pt} & \hspace{7pt} 4 \hspace{7pt} & \hspace{7pt} 5 \hspace{7pt} & \hspace{7pt} 6 \hspace{7pt}& \hspace{7pt} 7 \\
\hline
 H &  0.0 &  0.9 &  6.4 & 16.1 &  0.3 & {\bf 75.1} &  0.3 &  0.1 &  0.3 &  0.3 &  0.3 &  0.0  \\
Li &  0.2 &  1.8 &  4.7 &  2.6 &  0.1 & {\bf 81.7} &  1.8 &  4.1 &  1.9 &  0.9 &  0.1 &  0.0  \\
Be &  0.3 &  2.8 &  5.3 &  5.8 &  0.1 &  1.3 & {\bf 80.1} &  2.3 &  1.3 &  0.8 &  0.1 &       \\
 B &  0.2 &  4.9 & {\bf 22.9} & 10.1 &  0.6 &  2.4 &  1.8 & {\bf 56.5} &  0.2 &  0.2 &  0.1 &       \\
 C & {\bf 24.0} & 12.5 & {\bf 22.0} &  4.8 &  0.8 &  0.2 &  1.3 &  1.9 & {\bf 32.4} &      &      &       \\
 N &      & {\bf 83.9} &  3.2 &  1.3 &  0.3 &  0.3 &  0.1 &  0.4 &      & 10.5 &      &       \\
 O &      &  0.0 & {\bf 98.5} &  0.5 &  0.0 &  0.1 &  0.1 &  0.1 &  0.2 &  0.3 &  0.1 &  0.0  \\
 F &      &      &  0.2 & {\bf 99.6} &  0.0 &  0.1 &  0.0 &  0.0 &  0.1 &      &      &       \\
Na &  0.0 &  0.8 &  5.1 &  3.0 &  0.2 & {\bf 79.5} &  2.8 &  4.6 &  2.2 &  1.2 &  0.6 &  0.0  \\
Mg &  0.4 &  1.8 &  4.7 &  5.5 &  0.0 &  0.8 & {\bf 78.9} &  2.9 &  2.4 &  2.0 &  0.4 &       \\
Al &  0.4 &  0.9 & 13.5 &  5.4 &  0.0 &  2.3 &  1.5 & {\bf 74.3} &  1.0 &  0.6 &  0.2 &       \\
Si & 17.9 &  1.4 & 14.6 &  3.4 &  0.1 &  0.6 &  1.0 &  1.5 & {\bf 59.1} &  0.3 &      &       \\
 P &      & 12.1 & 15.1 &  3.4 &  0.7 &  1.5 &  2.0 &  1.1 &  3.6 & {\bf 60.5} &  0.0 &       \\
 S &      &  0.1 & {\bf 80.1} &  6.4 &  0.5 &  0.2 &  0.2 &  0.1 &  0.4 &  0.1 & 11.8 &       \\
Cl &      &      &  3.5 & {\bf 95.1} &  0.0 &  0.0 &  0.1 &  0.2 &  0.0 &  0.3 &      &  0.7  \\
 K &      &  0.4 &  2.7 &  4.0 &  0.2 & {\bf 81.8} &  2.5 &  4.2 &  1.8 &  1.6 &  0.7 &  0.0  \\
Ca &  0.3 &  3.2 &  4.1 &  1.9 &  0.1 &  0.6 & {\bf 80.6} &  4.9 &  2.6 &  1.1 &  0.4 &  0.1  \\
Sc &  1.9 &  1.6 & 12.8 &  7.1 &  0.1 & 11.9 & 10.5 & {\bf 49.5} &  2.8 &  1.2 &  0.6 &       \\
Ti &  0.6 &  1.2 & 19.1 &  2.9 &  0.1 &  0.6 &  6.8 & 12.1 & {\bf 54.9} &  1.3 &  0.3 &  0.0  \\
 V &  0.4 &  1.2 & 13.4 &  2.9 &  0.0 &  1.5 &  5.3 & {\bf 20.5} & 18.5 & {\bf 35.9} &  0.4 &  0.0  \\
Cr &  1.4 &  1.4 & {\bf 24.5} &  3.7 &  0.0 &  2.5 &  8.3 & {\bf 38.0} &  4.6 &  5.9 &  9.5 &       \\
Mn &  1.8 &  1.0 & 13.9 &  4.2 &  0.4 &  3.4 & {\bf 37.9} & 19.9 & 13.4 &  3.2 &  0.6 &  0.2  \\
Fe &  3.3 &  1.4 & 13.0 &  3.8 &  0.0 &  7.7 & {\bf 28.6} & {\bf 34.0} &  3.8 &  3.1 &  1.2 &       \\
Co &  5.8 &  1.8 & 11.4 &  2.8 &  0.1 & {\bf 24.3} & {\bf 35.6} & 10.8 &  6.4 &  0.9 &  0.2 &       \\
Ni &  0.7 &  5.6 & 14.7 &  5.4 &  0.0 & {\bf 21.7} & {\bf 39.4} &  7.0 &  4.0 &  1.2 &  0.3 &  0.0  \\
Cu &  0.4 &  2.0 & {\bf 26.4} &  7.6 &  0.1 & {\bf 31.7} & {\bf 25.3} &  3.9 &  1.5 &  0.8 &  0.3 &       \\
Zn &  0.0 &  1.5 & 10.2 &  3.0 &  0.1 &  1.1 & {\bf 78.6} &  2.7 &  1.7 &  0.9 &  0.2 &       \\
Ga &  0.3 &  2.1 & {\bf 20.7} &  2.3 &  0.1 &  4.4 &  5.2 & {\bf 63.4} &  0.9 &  0.5 &  0.1 &       \\
Ge & 16.6 &  1.2 & {\bf 30.9} &  0.6 &  0.0 &  0.5 &  5.2 &  3.9 & {\bf 40.6} &  0.4 &      &       \\
Se &      &  0.1 & {\bf 68.8} & 13.2 &  1.0 &  0.3 &  0.2 &  0.1 & 12.0 &  0.1 &  4.1 &       \\
Br &      &  0.1 &  2.1 & {\bf 95.1} &  0.4 &  0.1 &  0.5 &  0.5 &  0.0 &  1.2 &      &       \\
Rb &      &  0.2 &  3.3 & 10.6 &  0.5 & {\bf 74.5} &  2.4 &  4.8 &  1.8 &  1.1 &  0.8 &  0.0  \\
Sr &  0.1 &  1.8 &  8.1 &  1.3 &  0.1 &  0.8 & {\bf 76.9} &  5.0 &  3.6 &  1.8 &  0.5 &  0.1  \\
 Y &  0.4 &  0.5 & 15.3 &  3.3 &  0.3 &  4.1 &  1.6 & {\bf 68.6} &  3.0 &  2.1 &  0.8 &       \\
Zr &  1.2 &  2.2 &  5.8 & 11.8 &  0.2 & 11.1 & 13.0 & 11.6 & {\bf 42.4} &  0.5 &  0.2 &       \\
Nb &  0.4 &  1.3 & 15.1 &  3.7 &  0.0 &  5.1 & 10.9 & 14.0 &  6.4 & {\bf 42.8} &  0.2 &  0.1  \\
Mo &  0.5 &  1.2 & 18.6 &  2.9 &  0.0 &  0.3 & 10.5 &  4.2 &  5.6 & 11.6 & {\bf 44.7} &       \\
Ru &  1.7 &  1.9 & {\bf 28.8} &  2.1 &  0.2 &      & 15.3 &  5.9 & 12.5 & {\bf 27.6} &  4.1 &       \\
Rh &      &  0.6 & 16.0 &  4.5 &  0.1 & {\bf 33.1} &      & {\bf 31.1} & 14.2 &      &  0.4 &       \\
Pd &      &      & 12.9 & 11.8 &  0.1 &      & {\bf 58.5} & 10.0 &  6.0 &      &  0.7 &       \\
Ag &  0.1 &  0.4 & {\bf 22.4} &  4.1 &  0.2 & {\bf 60.1} &  3.7 &  2.5 &  2.2 &  2.2 &  2.0 &  0.1  \\
Cd &      &  0.9 & 11.1 &  4.2 &  0.1 &  0.5 & {\bf 78.0} &  2.2 &  1.5 &  0.9 &  0.5 &       \\
In &  0.1 &  0.3 & {\bf 21.9} &  3.4 &  0.2 & 12.6 &  7.1 & {\bf 51.6} &  1.6 &  1.0 &  0.3 &       \\
Sn &  0.3 &  1.1 & {\bf 22.5} &  7.5 &  0.1 &  0.6 & {\bf 29.0} &  2.3 & {\bf 36.1} &  0.5 &  0.0 &       \\
Sb &      & 19.9 & {\bf 30.3} &  8.4 &  0.5 &  0.2 &  1.2 & {\bf 23.8} &  0.3 & 15.0 &  0.4 &       \\
Te &      &  0.0 & {\bf 52.5} & 16.1 &  0.5 &  0.8 &  0.9 &  0.3 & {\bf 23.8} &  0.3 &  4.7 &       \\
 I &      &      &  4.4 & {\bf 85.1} &  0.1 &  0.6 &  0.0 &  0.3 &  0.0 &  8.7 &      &  0.8  \\


Cs &      &  0.3 &  2.1 & 13.1 &  0.3 & {\bf 70.9} &  3.1 &  6.9 &  1.8 &  1.0 &  0.6 &  0.0  \\
Ba &  0.1 &  1.2 &  6.8 &  1.3 &  0.0 &  1.0 & {\bf 78.4} &  5.7 &  2.3 &  2.4 &  0.7 &  0.1  \\
La &  0.6 &  1.4 &  8.5 &  2.4 &  0.1 &  4.2 &  8.7 & {\bf 69.5} &  2.3 &  1.5 &  0.9 &  0.0  \\
Ce &  1.9 &  1.7 &  8.2 &  2.8 &  0.1 &  1.4 & 14.2 & {\bf 52.3} & 16.3 &  0.8 &  0.2 &  0.1  \\
Pr &  0.6 &  1.2 &  8.2 &  2.7 &  0.4 &  1.3 &  6.9 & {\bf 72.6} &  3.7 &  1.4 &  1.0 &       \\
Nd &  0.4 &  0.8 & 10.0 &  1.1 &  0.1 &  1.3 &  9.2 & {\bf 70.8} &  3.0 &  1.8 &  1.6 &       \\
Sm &  0.3 &  1.0 & 12.0 &  1.1 &  0.2 &  1.7 & 10.3 & {\bf 67.2} &  2.7 &  2.1 &  1.4 &  0.1  \\
Eu &  0.3 &  2.9 & 12.2 &  3.5 &  0.1 &  1.3 & {\bf 43.4} & {\bf 30.1} &  2.9 &  2.0 &  1.3 &       \\
Gd &  0.2 &  0.8 & 12.1 &  2.4 &  0.1 &  2.3 &  4.8 & {\bf 70.1} &  3.0 &  2.9 &  1.2 &       \\
Tb &  1.6 &  1.0 &  9.2 &  3.0 &  0.4 &  9.1 & 12.7 & {\bf 52.0} &  8.9 &  1.4 &  0.7 &       \\
Dy &  0.7 &  1.1 & 12.0 &  1.7 &  0.2 &  2.4 &  9.4 & {\bf 67.7} &  2.6 &  1.9 &  0.4 &       \\
Ho &  1.1 &  0.6 & 13.6 &  3.0 &  0.2 &  2.5 & 10.5 & {\bf 64.2} &  2.0 &  1.5 &  0.8 &  0.2  \\
Er &  0.9 &  0.4 & 12.1 &  3.0 &  0.3 &  3.7 &  2.2 & {\bf 72.2} &  2.3 &  1.8 &  1.0 &  0.2  \\
Tm &  2.0 &  0.7 & 17.1 &  2.1 &  0.2 &  1.9 & 11.7 & {\bf 59.8} &  2.3 &  1.1 &  1.1 &       \\
Lu &  0.4 &  0.4 & 16.1 &  2.5 &  0.3 &  3.6 &  1.4 & {\bf 70.4} &  2.8 &  1.0 &  0.9 &  0.3  \\
Hf &  0.8 &  2.0 & 10.8 &  6.0 &  0.1 &  2.2 &  7.7 &  4.5 & {\bf 65.2} &  0.6 &  0.2 &       \\
Ta &  0.5 &  2.1 & 16.8 &  2.1 &  1.3 & 10.1 &  7.4 &  7.0 &  6.9 & {\bf 45.8} &  0.2 &       \\
 W &      &  2.8 & {\bf 29.4} &  3.5 &  0.0 &  0.1 & 10.3 &  3.0 &  3.2 &  7.5 & {\bf 40.2} &       \\
Re &  2.5 &  5.2 & {\bf 20.8} &  4.5 &  0.3 &  1.1 &  1.2 & {\bf 21.7} &  5.2 & 11.4 &  6.9 & 19.2  \\
Os &      &  3.8 & 14.6 &  3.8 &  1.3 &      &      &      & {\bf 76.6} &      &      &       \\
Ir &      &  1.4 & {\bf 30.0} &  3.8 &  0.1 &      &  1.4 & {\bf 20.9} & 17.4 & {\bf 20.9} &  4.0 &       \\
Pt &      &      & {\bf 27.6} &  9.3 &  0.2 &  0.2 & {\bf 28.9} &  5.8 & {\bf 28.1} &      &      &       \\
Au &      &      & 13.3 & 13.3 &  0.5 &      &      & {\bf 68.3} &  0.6 &  1.6 &  2.4 &       \\
Hg &      &      & 15.5 &  8.8 &  2.6 & 12.2 & {\bf 57.8} &  0.4 &  1.0 &  0.7 &  0.9 &       \\
Tl &      &      & 13.2 &  7.6 &  0.2 & {\bf 61.5} &  1.9 & 10.3 &  2.0 &  1.6 &  1.7 &  0.0  \\
Pb &      &      & {\bf 23.2} &  5.2 &  0.0 &      & {\bf 67.5} &  0.6 &  2.7 &  0.4 &  0.3 &       \\
Bi &      &  3.9 & {\bf 21.9} &  3.6 &  0.3 &  1.0 &  2.9 & {\bf 57.5} &  2.2 &  5.6 &  1.2 &       \\
\hline
 \hspace{20pt}  & \hspace{7pt}  -4 \hspace{7pt} & \hspace{7pt} -3 \hspace{7pt} & \hspace{7pt} -2 \hspace{7pt} & \hspace{7pt} -1 \hspace{7pt} & \hspace{7pt} 0 \hspace{7pt} & \hspace{7pt} 1 \hspace{7pt} & \hspace{7pt} 2 \hspace{7pt} & \hspace{7pt} 3 \hspace{7pt} & \hspace{7pt} 4 \hspace{7pt} & \hspace{7pt} 5 \hspace{7pt} & \hspace{7pt} 6 \hspace{7pt}& \hspace{7pt} 7 \\
\hline
\caption{\label{tab:oxi}Distribution of the oxidation states for elements from Cs to Bi. Oxidation states with ratio over 20\% are shown in bold.}
\end{longtable}

The average vacancy formation energies of element at given oxidation state are listed in Table \ref{tab:oxi_2}.

\begin{longtable}{lccccccccccccc}
\hline
 \hspace{20pt}  & \hspace{7pt}  -4 \hspace{7pt} & \hspace{7pt} -3 \hspace{7pt} & \hspace{7pt} -2 \hspace{7pt} & \hspace{7pt} -1 \hspace{7pt} & \hspace{7pt} 0 \hspace{7pt} & \hspace{7pt} 1 \hspace{7pt} & \hspace{7pt} 2 \hspace{7pt} & \hspace{7pt} 3 \hspace{7pt} & \hspace{7pt} 4 \hspace{7pt} & \hspace{7pt} 5 \hspace{7pt} & \hspace{7pt} 6 \hspace{7pt}& \hspace{7pt} 7 \\
\hline
 H & 2.4 & 2.4 & 2.6 & 1.3 & 2.1 & {\bf 2.7} & 2.0 & 2.6 & 2.7 & 2.8 & 2.7 & 3.4  \\
Li & 0.9 & 1.7 & 3.4 & 3.5 & 0.6 & {\bf 3.5} & 2.9 & 3.6 & 3.6 & 3.7 & 4.2 & 3.5  \\
Be & 1.4 & 3.6 & 8.0 & 5.0 & 1.0 & 6.7 & {\bf 8.0} & 8.2 & 6.7 & 8.8 & 7.9 &       \\
 B & 4.7 & 2.1 & {\bf 5.6} & 3.0 & 2.4 & 4.6 & 4.8 & {\bf 9.7} & 2.2 & 10.4 & 7.6 &       \\
 C & {\bf 2.1} & 3.0 & {\bf 5.1} & 3.1 & 7.1 & 9.2 & 5.5 & 7.7 & {\bf 8.5} &      &      &       \\
 N &      & {\bf 3.7} & 4.8 & 3.4 & 5.5 & 4.1 & -1.0 & 3.9 &      & 7.2 &      &       \\
 O &      & 2.8 & {\bf 4.4} & 3.4 & 2.7 & 4.8 & 4.7 & 4.3 & 4.0 & 4.3 & 3.2 & 3.0  \\
 F &      &      & 4.1 & {\bf 3.8} & 0.9 & 3.9 & 3.1 & 5.1 & 3.6 &      &      &       \\
Na & 1.0 & 1.2 & 2.6 & 3.4 & 0.4 & {\bf 3.1} & 2.9 & 3.1 & 2.9 & 3.3 & 3.2 & 4.1  \\
Mg & 1.2 & 3.4 & 5.8 & 7.0 & 0.8 & 7.0 & {\bf 6.6} & 6.3 & 7.5 & 7.0 & 8.5 &       \\
Al & 2.2 & 3.3 & 9.7 & 11.4 & 0.6 & 10.1 & 9.6 & {\bf 9.9} & 11.0 & 11.9 & 12.1 &       \\
Si & 2.6 & 8.5 & 11.2 & 3.4 & 2.6 & 11.6 & 4.1 & 3.7 & {\bf 12.6} & 12.7 &      &       \\
 P &      & 3.1 & 10.7 & 3.1 & 1.0 & 9.4 & 5.5 & 5.4 & 4.9 & {\bf 13.7} & 9.3 &       \\
 S &      & 2.5 & {\bf 2.6} & 1.3 & 1.2 & 6.6 & 5.0 & 10.6 & 7.2 & 12.4 & 13.4 &       \\
Cl &      &      & 3.4 & {\bf 2.2} & 0.8 & 2.0 & 2.6 & 3.7 & 1.8 & 6.1 &      & 8.2  \\
 K &      & 1.2 & 2.5 & 3.7 & 0.7 & {\bf 3.1} & 3.0 & 3.1 & 3.2 & 3.5 & 3.7 & 4.0  \\
Ca & 2.6 & 3.6 & 6.8 & 6.4 & 1.1 & 5.8 & {\bf 6.9} & 7.8 & 7.6 & 6.8 & 8.4 & 6.2  \\
Sc & 3.2 & 3.5 & 9.4 & 7.9 & 1.6 & 3.4 & 3.0 & {\bf 9.3} & 11.1 & 10.2 & 12.5 &       \\
Ti & 3.7 & 4.6 & 10.0 & 7.8 & 1.6 & 6.6 & 4.4 & 5.4 & {\bf 11.8} & 12.3 & 14.1 & 13.4  \\
 V & 1.9 & 5.9 & 7.6 & 8.6 & 3.0 & 7.1 & 4.1 & {\bf 5.4} & 7.7 & {\bf 11.0} & 7.4 & 7.6  \\
Cr & 2.6 & 6.4 & {\bf 4.6} & 7.9 & 3.0 & 5.3 & 3.1 & {\bf 5.2} & 6.5 & 7.4 & 10.5 &       \\
Mn & 2.0 & 4.0 & 4.0 & 5.7 & 1.6 & 3.0 & {\bf 3.1} & 5.0 & 5.8 & 5.1 & 5.9 & 8.1  \\
Fe & 1.7 & 3.0 & 2.7 & 6.4 & 1.9 & 1.8 & {\bf 1.5} & {\bf 4.7} & 3.7 & 2.8 & 3.6 &       \\
Co & 1.5 & 1.7 & 2.0 & 4.4 & 1.3 & {\bf 1.4} & {\bf 1.8} & 2.7 & 2.9 & 2.7 & 1.8 &       \\
Ni & 0.9 & 1.4 & 1.4 & 3.1 & 0.9 & {\bf 1.3} & {\bf 1.6} & 2.0 & 1.7 & 2.4 & 1.6 & 1.7  \\
Cu & 0.9 & 0.4 & {\bf 1.0} & 2.9 & 1.0 & {\bf 0.6} & {\bf 2.1} & 3.5 & 1.3 & 3.6 & 1.7 &       \\
Zn & 0.7 & 1.9 & 3.2 & 5.5 & 0.3 & 3.6 & {\bf 3.8} & 3.8 & 4.6 & 4.6 & 5.2 &       \\
Ga & 1.0 & 4.2 & {\bf 5.7} & 10.5 & 0.6 & 3.7 & 2.8 & {\bf 6.8} & 8.7 & 9.3 & 8.4 &       \\
Ge & 2.4 & 4.7 & {\bf 5.1} & 4.9 & 1.9 & 7.5 & 2.7 & 3.2 & {\bf 8.8} & 10.1 &      &       \\
Se &      & 3.1 & {\bf 2.2} & 1.2 & 1.0 & 2.4 & 7.1 & 8.3 & 7.5 & 9.1 & 11.1 &       \\
Br &      & 2.6 & 2.8 & {\bf 2.0} & 0.7 & 0.1 & 1.9 & 4.2 & 1.5 & 5.5 &      &       \\
Rb &      & 1.1 & 2.4 & 3.6 & 1.0 & {\bf 3.1} & 2.8 & 3.2 & 3.2 & 3.5 & 3.9 & 4.2  \\
Sr & 2.7 & 3.3 & 6.5 & 6.2 & 0.8 & 5.0 & {\bf 6.5} & 7.0 & 7.3 & 6.8 & 6.1 & 5.1  \\
 Y & 3.0 & 5.9 & 10.5 & 11.4 & 1.8 & 4.8 & 7.1 & {\bf 10.3} & 12.0 & 11.7 & 10.3 &       \\
Zr & 6.2 & 4.8 & 11.6 & 9.7 & 2.2 & 3.0 & 4.2 & 4.0 & {\bf 12.8} & 15.6 & 15.1 &       \\
Nb & 3.3 & 6.3 & 10.9 & 7.5 & 3.3 & 3.7 & 3.6 & 4.1 & 5.7 & {\bf 12.6} & 15.6 & 15.9  \\
Mo & 4.2 & 9.5 & 9.5 & 6.0 & 3.8 & 10.4 & 2.9 & 3.8 & 4.5 & 7.5 & {\bf 10.7} &       \\
Ru & 1.7 & 2.1 & {\bf 5.7} & 10.3 & 1.8 &      & 1.9 & 2.3 & 4.9 & {\bf 7.4} & 9.6 &       \\
Rh &      & 2.0 & 3.1 & 7.4 & 1.1 & {\bf 1.6} &      & {\bf 2.6} & 3.9 &      & 3.7 &       \\
Pd &      &      & 2.4 & 3.8 & 1.1 &      & {\bf 1.7} & 1.6 & 3.1 &      & 2.3 &       \\
Ag & 0.6 & 0.4 & {\bf 0.5} & 1.6 & 0.6 & {\bf 0.5} & 2.1 & 1.8 & 0.6 & 0.6 & 0.5 & 0.3  \\
Cd &      & 1.4 & 3.0 & 4.2 & 0.3 & 3.1 & {\bf 3.1} & 3.8 & 4.0 & 4.9 & 4.4 &       \\
In & 1.1 & 2.6 & {\bf 4.1} & 4.1 & 0.4 & 1.9 & 2.7 & {\bf 5.3} & 6.6 & 7.9 & 7.4 &       \\
Sn & 0.8 & 2.7 & {\bf 5.8} & 5.9 & 1.0 & 9.3 & {\bf 3.3} & 4.7 & {\bf 7.2} & 8.9 & 10.6 &       \\
Sb &      & 2.1 & {\bf 4.7} & 3.4 & 6.0 & 14.6 & 10.0 & {\bf 4.6} & 9.8 & 12.6 & 8.2 &       \\
Te &      & 2.6 & {\bf 3.4} & 1.7 & 1.1 & 5.4 & 6.2 & 3.1 & {\bf 7.6} & 8.8 & 12.6 &       \\
 I &      &      & 7.5 & {\bf 1.5} & 0.5 & 1.0 & 6.9 & 3.5 & 10.1 & 8.1 &      & 11.2  \\
Cs &      & 0.8 & 2.3 & 3.5 & 0.5 & {\bf 3.1} & 3.2 & 3.2 & 3.2 & 3.6 & 3.5 & 4.4  \\
Ba & 2.3 & 2.9 & 6.2 & 5.1 & 0.8 & 5.4 & {\bf 6.4} & 6.6 & 7.1 & 6.8 & 7.2 & 6.8  \\
La & 3.4 & 4.0 & 7.8 & 5.4 & 1.4 & 3.3 & 4.0 & {\bf 8.8} & 10.5 & 10.1 & 11.3 & 4.6  \\
Ce & 3.3 & 3.6 & 6.9 & 5.9 & 1.3 & 4.2 & 2.9 & {\bf 5.4} & 8.9 & 10.0 & 6.8 & 14.6  \\
Pr & 3.5 & 4.6 & 8.4 & 5.1 & 1.5 & 4.7 & 4.2 & {\bf 8.3} & 9.1 & 10.3 & 12.4 &       \\
Nd & 3.1 & 4.6 & 9.0 & 6.0 & 1.6 & 5.6 & 4.5 & {\bf 9.4} & 10.6 & 12.0 & 12.2 &       \\
Sm & 3.2 & 4.1 & 8.3 & 6.5 & 1.5 & 5.8 & 4.1 & {\bf 9.2} & 10.4 & 12.0 & 12.1 & 13.5  \\
Eu & 2.6 & 3.5 & 7.3 & 5.7 & 1.0 & 7.9 & {\bf 4.4} & {\bf 9.0} & 8.2 & 9.1 & 10.2 &       \\
Gd & 3.3 & 4.1 & 9.5 & 7.5 & 1.4 & 5.4 & 5.1 & {\bf 10.1} & 12.5 & 11.8 & 12.4 &       \\
Tb & 3.4 & 3.9 & 8.4 & 7.9 & 1.9 & 3.5 & 3.2 & {\bf 9.2} & 6.6 & 10.6 & 12.7 &       \\
Dy & 2.8 & 3.3 & 9.3 & 10.8 & 1.8 & 5.1 & 4.0 & {\bf 9.0} & 12.5 & 12.1 & 12.6 &       \\
Ho & 3.0 & 3.4 & 9.2 & 13.0 & 1.8 & 5.4 & 3.9 & {\bf 8.9} & 12.5 & 11.9 & 12.6 & 13.9  \\
Er & 3.3 & 4.6 & 9.3 & 11.3 & 1.6 & 5.5 & 6.4 & {\bf 9.0} & 12.6 & 11.3 & 12.6 & 11.8  \\
Tm & 3.3 & 3.1 & 9.2 & 13.6 & 1.6 & 5.3 & 4.1 & {\bf 9.0} & 11.9 & 11.8 & 12.6 &       \\
Lu & 4.7 & 6.5 & 10.0 & 13.7 & 1.5 & 7.9 & 6.5 & {\bf 9.5} & 12.8 & 11.9 & 13.0 & 13.8  \\
Hf & 4.9 & 6.8 & 14.1 & 15.8 & 2.2 & 7.9 & 4.9 & 4.4 & {\bf 9.9} & 16.7 & 16.5 &       \\
Ta & 3.8 & 8.9 & 13.1 & 10.3 & 3.7 & 3.6 & 4.0 & 4.4 & 5.2 & {\bf 13.6} & 16.8 &       \\
 W &      & 10.5 & {\bf 9.7} & 4.7 & 3.6 & 9.7 & 1.8 & 2.7 & 3.9 & 6.2 & {\bf 9.3} &       \\
Re & 2.3 & 5.6 & {\bf 6.8} & 4.5 & 1.3 & 9.0 & 6.9 & {\bf 3.5} & 4.7 & 3.4 & 9.1 & 15.1  \\
Os &      & 4.1 & 5.5 & 5.3 & 2.6 &      &      &      & {\bf 3.1} &      &      &       \\
Ir &      & 4.3 & {\bf 5.1} & 4.4 & 0.8 &      & 5.8 & {\bf 2.6} & 4.4 & {\bf 5.9} & 8.9 &       \\
Pt &      &      & {\bf 4.6} & 5.5 & 0.9 & 2.8 & {\bf 2.2} & 3.3 & {\bf 5.7} &      &      &       \\
Au &      &      & 2.9 & 4.9 & 0.5 &      &      & {\bf 3.3} & 3.1 & 4.3 & 3.5 &       \\
Hg &      &      & 1.7 & 2.1 & 0.0 & 1.8 & {\bf 1.9} & 1.9 & 1.7 & 2.2 & 2.0 &       \\
Tl &      &      & 1.8 & 2.1 & 0.3 & {\bf 1.3} & 1.6 & 4.1 & 1.4 & 2.0 & 1.8 & 2.2  \\
Pb &      &      & {\bf 3.8} & 4.9 & 0.2 &      & {\bf 3.8} & 5.7 & 6.9 & 4.1 & 4.5 &       \\
Bi &      & 2.2 & {\bf 5.2} & 6.8 & 0.6 & 1.4 & 2.2 & {\bf 4.9} & 5.3 & 9.8 & 6.5 &       \\
\hline
 \hspace{20pt}  & \hspace{7pt}  -4 \hspace{7pt} & \hspace{7pt} -3 \hspace{7pt} & \hspace{7pt} -2 \hspace{7pt} & \hspace{7pt} -1 \hspace{7pt} & \hspace{7pt} 0 \hspace{7pt} & \hspace{7pt} 1 \hspace{7pt} & \hspace{7pt} 2 \hspace{7pt} & \hspace{7pt} 3 \hspace{7pt} & \hspace{7pt} 4 \hspace{7pt} & \hspace{7pt} 5 \hspace{7pt} & \hspace{7pt} 6 \hspace{7pt}& \hspace{7pt} 7 \\
\hline
\caption{\label{tab:oxi_2}Distribution of the average vacancy formation energies for elements from Cs to Bi. Oxidation states with ratio over 20\% are shown in bold (same as in Table \ref{tab:oxi}).}
\end{longtable}

\begin{figure*}[h]
    \centering
    \includegraphics[width=0.51\linewidth]{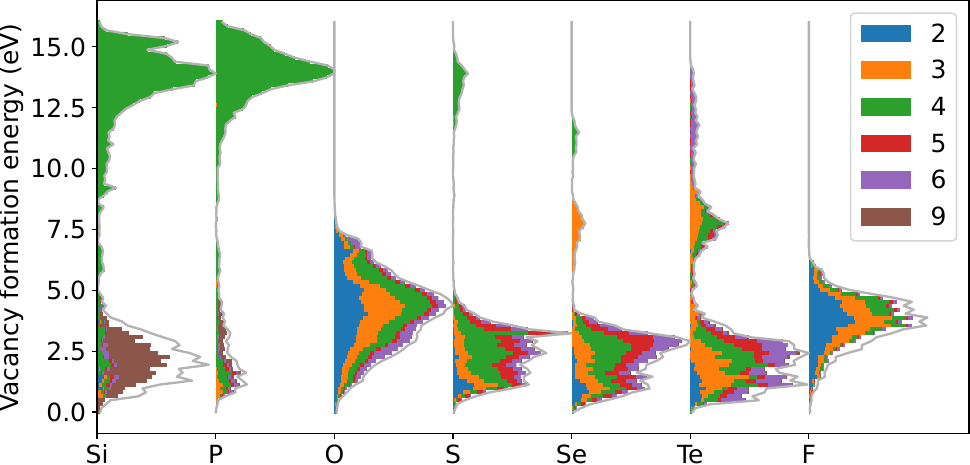}
    \caption{Vacancy formation energies of various elements with different coordination numbers.}
    \label{fig:Coord_others}
\end{figure*}

\begin{figure*}[h]
    \centering
    \includegraphics[width=0.51\linewidth]{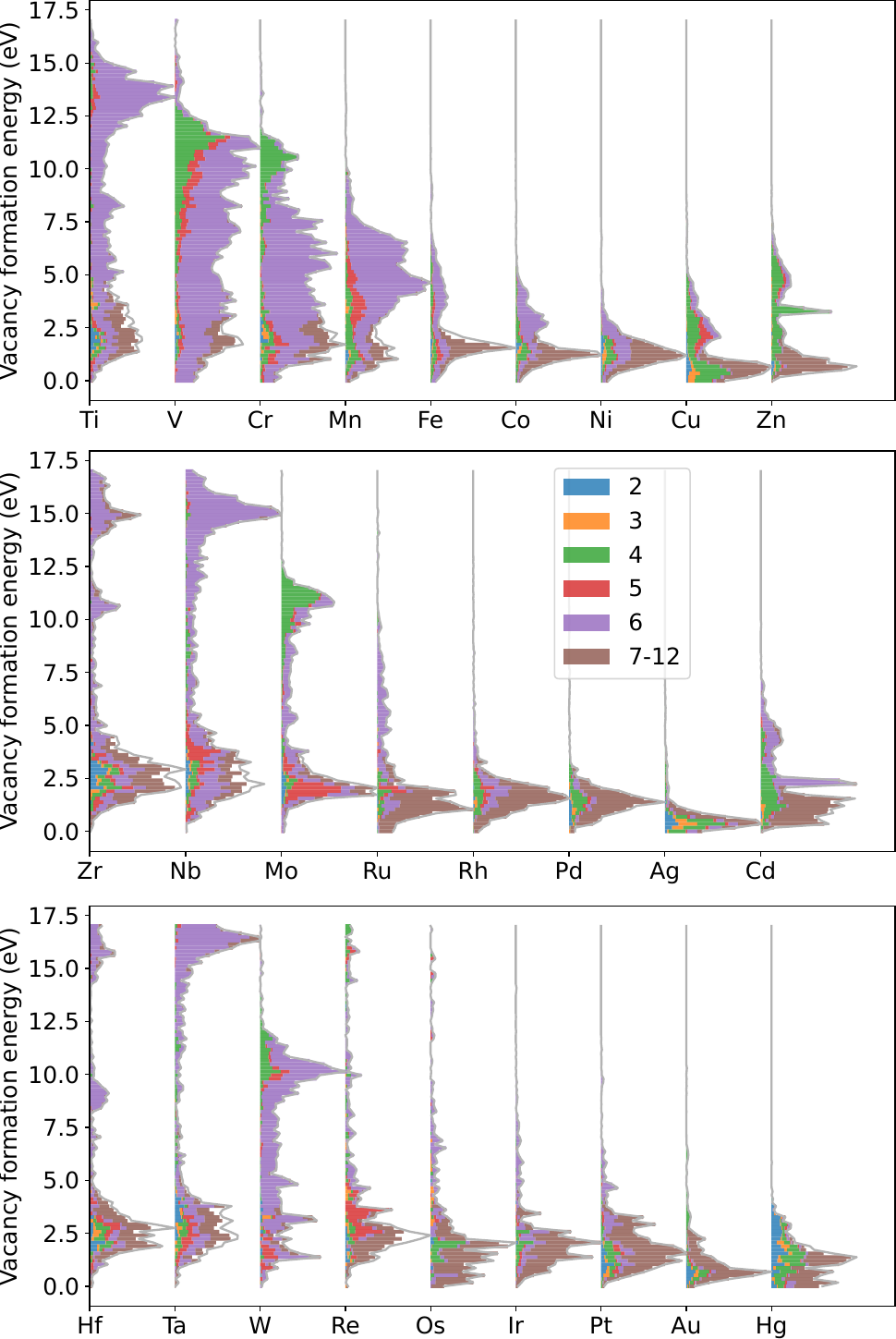}
    \caption{Vacancy formation energies of transition metals with different coordination numbers.}
    \label{fig:Coord_TM}
\end{figure*}

In addition to the oxidation number, the main text also discusses the impact of the coordination number on the vacancy formation energies. Results can be found in Fig.\ \ref{fig:Coord_TM} for transition metals and in Fig.\ \ref{fig:Coord_others} for other elements.

\begin{figure*}[h]
    \centering
    \includegraphics[width=\linewidth]{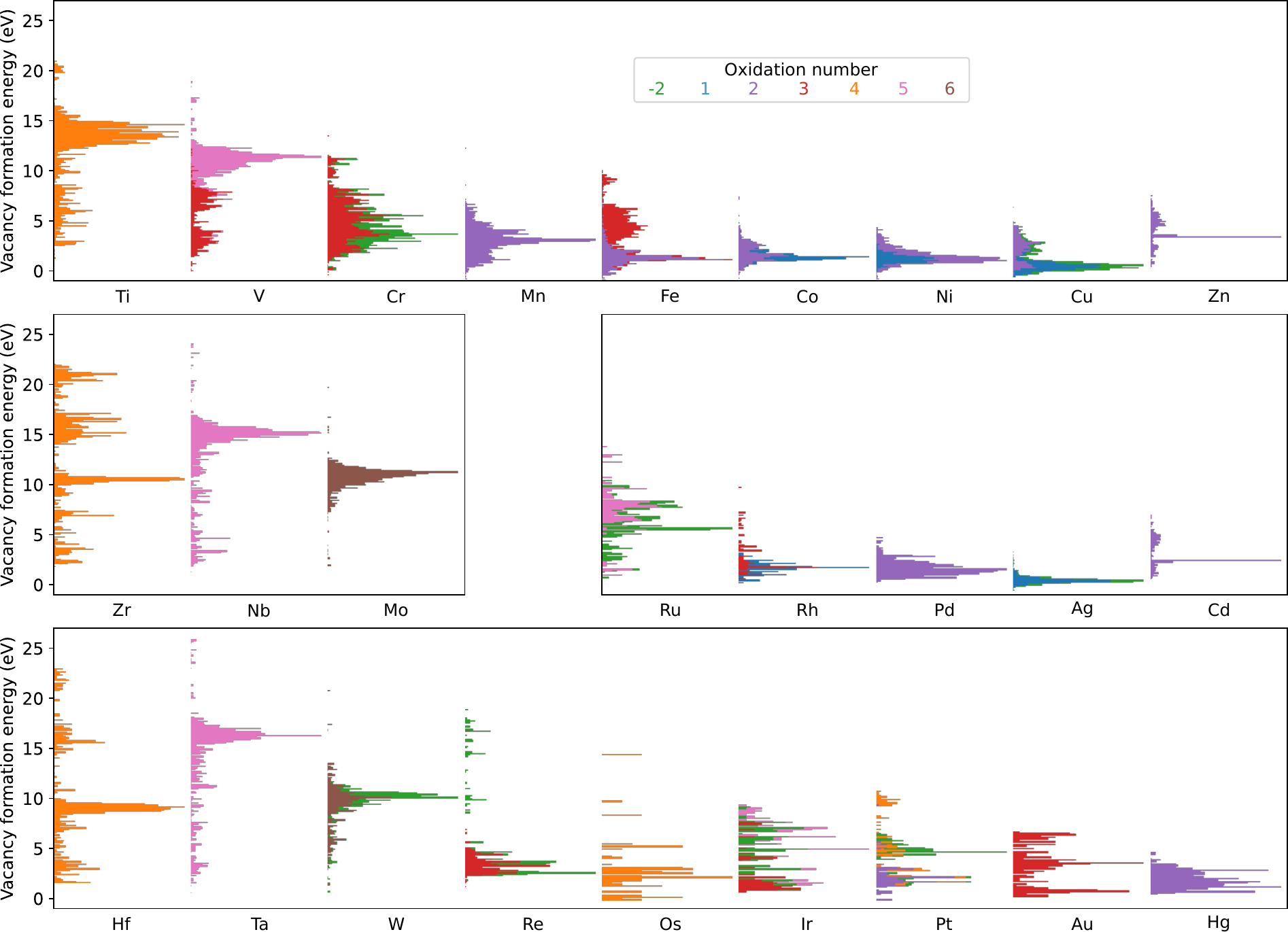}
    \caption{Stacked histograms of the vacancy formation energies for the 3d, 4d, and 5d transition metals.}
    \label{fig:Oxi_SI}
\end{figure*}

Fig.\ 2(d) of the main text shows the vacancy formation energy for various oxidation states of 3d transition metals. A complete picture including all transition metal elements is shown in Fig.\ \ref{fig:Oxi_SI}.

\vspace{2cm}

\newpage

\newpage

\newpage

\section{Best candidates}

\begin{longtable}{lcccccc}
\hline
MP ID      & Etched & Layer     & HEE & LNE & LDFE & Inital Dim. \\ 
\hline 
mp-1232339  & Li  & \ce{C}  & -3.28  &  6.39  &  6.67  & 2D : 0.499   \\ 
mp-1021323  & Li  & \ce{C}  &  -3.3  &   6.4  &  6.67  & 02D : 0.494   \\ 
mp-1001581  & Li  & \ce{C}  & -3.19  &  5.53  &  6.67  & 3D : 0.54   \\ 
mp-28930  & K  & \ce{C}  & -3.18  &  5.35  &  6.67  & 3D : 0.803   \\ 
mp-568643  & Rb  & \ce{C}  &  -3.2  &  5.28  &  6.67  & 3D : 0.847   \\ 
mp-28861  & Cs  & \ce{C}  & -3.26  &  5.51  &  6.67  & 3D : 0.922   \\ 
mp-1208630  & Sr  & \ce{C}  & -5.61  &  4.07  &  6.67  & 3D : 0.907   \\ 
mp-1214417  & Ba  & \ce{C}  & -5.57  &   4.2  &  6.68  & 3D : 0.926   \\ 
mp-1025297  & Sm  & \ce{C}  & -6.48  &  4.42  &  6.67  & 3D : 1.0   \\ 
mp-1103990  & Eu  & \ce{C}  & -4.68  &  3.71  &  6.68  & 3D : 0.984   \\ 

\hline 
mp-19755  & Li  & \ce{TiS2}  & -1.46  &  1.39  &  2.16  & 2D : 0.742   \\ 
mp-1223278  & Li  & \ce{TiS2}  & -1.21  &  1.39  &  2.16  & 3D : 0.734   \\
mp-675056  & Na  & \ce{TiS2}  & -1.04  &  1.92  &  2.16  & 2D : 0.899   \\
mp-1048589  & Ca  & \ce{TiS2}  & -1.86  &  2.08  &  2.16  & 2D : 0.96   \\ 
mp-1048662  & Ca  & \ce{TiS2}  &  -2.0  &  1.63  &  2.16  & 3D : 1.0   \\  
mp-1048666  & Ca  & \ce{TiS2}  & -2.11  & 0.837  &  2.16  & 3D : 1.0   \\ 
mp-1386933  & Ca  & \ce{TiS2}  & -1.96  &  1.33  &  2.16  & 3D : 1.0   \\ 
mp-756195  & Li  & \ce{VS2}  &  -2.2  &  1.87  &  1.45  & 3D : 1.0   \\ 
mp-1177739  & Li  & \ce{VS2}  & -1.52  &  1.52  &  1.39  & 3D : 0.931   \\
mp-676586  & Na  & \ce{VS2}  & -1.09  &  1.59  &  1.39  & 3D : 0.911   \\
mp-675593  & K  & \ce{VS2}  & -1.44  &  1.57  &  1.55  & 3D : 0.994   \\  
mp-1223993  & K  & \ce{VS2}  & -1.37  &  1.78  &  1.39  & 3D : 0.962   \\ 
mp-1400988  & Ca  & \ce{VS2}  & -2.39  &  1.46  &   1.6  & 3D : 1.0   \\    
mp-1216498  & V  & \ce{VS2}  & -0.893  & 0.882  &  1.53  & 3D : 1.0   \\   
mp-4226  & Li  & \ce{CrS2}  & -1.05  &  1.62  &  1.18  & 3D : 0.617   \\ 
mp-755947  & Li  & \ce{CrS2}  & -0.935  &  2.03  &  1.15  & 3D : 0.838   \\
mp-1238813  & Li  & \ce{CrS2}  & -1.21  &  1.61  &  1.15  & 3D : 0.725   \\
mp-1238817  & Li  & \ce{CrS2}  & -0.753  &  1.67  &  1.18  & 3D : 0.674   \\
mp-5693  & Na  & \ce{CrS2}  & -1.16  &  1.87  &  1.18  & 3D : 0.93   \\  
mp-637292  & Na  & \ce{CrS2}  & -1.15  &  1.88  &  1.18  & 3D : 0.905   \\
mp-1238847  & Na  & \ce{CrS2}  & -0.869  &   1.4  &  1.15  & 3D : 0.866   \\ 
mp-4026  & K  & \ce{CrS2}  & -1.74  &  2.02  &  1.18  & 3D : 0.997   \\  
mp-1238855  & K  & \ce{CrS2}  & -1.56  &  1.37  &  1.15  & 3D : 0.971   \\
mp-1238878  & Rb  & \ce{CrS2}  & -1.13  &  1.32  &  1.15  & 3D : 0.986   \\
mp-1238845  & Cs  & \ce{CrS2}  & -0.878  &  1.33  &  1.15  & 3D : 1.0   \\ 
mp-1390152  & Mg  & \ce{CrS2}  & -1.63  &  1.57  &  1.18  & 3D : 0.917   \\ 
mp-2227138  & Mg  & \ce{CrS2}  & -2.55  &  1.39  &  1.18  & 2D : 0.889   \\ 
mp-1393946  & Ca  & \ce{CrS2}  & -0.965  &  1.45  &  1.18  & 3D : 1.0   \\
mp-1221432  & Na  & \ce{CrSe2}  & -1.11  &  1.34  &  1.39  & 3D : 0.944   \\
mp-30248  & Li  & \ce{MoS2}  & -1.61  &  2.88  &  1.66  & 3D : 0.682   \\  
mp-1025173  & Ti  & \ce{MoS2}  & -1.97  &  1.42  &  1.63  & 3D : 1.0   \\  
mp-555370  & V  & \ce{MoS2}  & -1.71  &  1.78  &  1.63  & 3D : 0.999   \\ 
mp-1192730  & V  & \ce{MoS2}  & -1.75  &  1.76  &  1.62  & 3D : 0.999   \\ 
mp-542188  & Cr  & \ce{MoS2}  & -2.06  &  1.68  &  1.62  & 3D : 0.991   \\
mp-1087488  & In  & \ce{MoS2}  & -1.71  &  3.18  &  2.87  & 2D : 0.474   \\ 
mp-1199798  & Cs  & \ce{ReS2}  & -1.16  &  1.73  & 0.752  & 3D : 1.0   \\  
mp-1077996  & Cr  & \ce{RhSe2}  & -1.83  &  1.01  &  1.01  & 3D : 1.0   \\
mp-1078162  & Fe  & \ce{RhSe2}  & -1.35  &  1.04  &  1.01  & 3D : 1.0   \\ 
mp-1078249  & Co  & \ce{RhSe2}  & -0.896  & 0.996  &  1.01  & 3D : 0.999   \\ 
mp-1077939  & Cr  & \ce{RhTe2}  & -2.29  &  1.31  &  1.68  & 3D : 0.999   \\ 
\hline 
mp-7936  & Li  & \ce{NbS2}  & -1.56  &   3.2  &  1.78  & 3D : 0.719   \\ 
mp-767218  & Li  & \ce{NbS2}  & -1.14  &  2.58  &  1.82  & 3D : 0.695   \\
mp-7937  & Na  & \ce{NbS2}  &  -1.8  &  3.63  &  1.78  & 3D : 0.955   \\    
mp-1221395  & Na  & \ce{NbS2}  &  -1.6  &  2.64  &  1.78  & 2D : 0.943   \\
mp-1221460  & Na  & \ce{NbS2}  & -1.15  &  2.84  &  1.78  & 3D : 0.906   \\
mp-1224044  & K  & \ce{NbS2}  & -1.48  &  2.89  &  1.82  & 3D : 0.991   \\ 
mp-7938  & K  & \ce{NbS2}  & -2.66  &  3.73  &  1.82  & 3D : 1.0   \\ 
mp-1229211  & Cs  & \ce{NbS2}  & -1.41  &  2.62  &  1.79  & 3D : 1.0   \\ 
mp-1188929  & Ti  & \ce{NbS2}  & -0.897  &  3.18  &  1.82  & 3D : 1.0   \\  
mp-1189260  & Cr  & \ce{NbS2}  & -0.984  &   2.9  &  1.82  & 3D : 1.0   \\ 
mp-15958  & V  & \ce{NbS2}  & -0.815  &  3.01  &  1.82  & 3D : 1.0   \\ 
mp-10199  & Mn  & \ce{NbS2}  & -1.58  &  2.74  &  1.78  & 3D : 0.959   \\ 
mp-1220734  & Fe  & \ce{NbS2}  & -2.16  &  2.33  &  1.78  & 2D : 0.952   \\ 
mp-20621  & In  & \ce{NbS2}  &  -1.5  &  2.87  &  1.82  & 02D : 0.614   \\ 
mp-1018022  & In  & \ce{NbS2}  & -0.902  &  2.88  &  1.82  & 3D : 0.546   \\
mp-1025496  & Li  & \ce{NbSe2}  & -1.92  &  2.61  &  2.05  & 3D : 0.797   \\
mp-7939  & Na  & \ce{NbSe2}  & -1.94  &  2.84  &  2.11  & 3D : 0.985   \\ 
mp-1221396  & Na  & \ce{NbSe2}  & -1.84  &  2.85  &  2.05  & 2D : 0.944   \\ 
mp-1221482  & Na  & \ce{NbSe2}  & -1.84  &  2.84  &  2.05  & 2D : 0.913   \\
mp-7940  & K  & \ce{NbSe2}  & -2.59  &  2.66  &  2.11  & 3D : 1.0   \\  
mp-1224047  & K  & \ce{NbSe2}  &  -1.7  &  2.97  &  2.11  & 3D : 1.0   \\ 
mp-1025291  & Ti  & \ce{NbSe2}  & -2.17  &  1.97  &  2.26  & 3D : 0.994   \\
mp-1188631  & Ti  & \ce{NbSe2}  & -2.11  &  2.47  &  2.04  & 3D : 1.0   \\ 
mp-1025195  & V  & \ce{NbSe2}  & -2.19  &  2.19  &  2.26  & 3D : 0.99   \\
mp-1105840  & V  & \ce{NbSe2}  & -2.08  &  2.66  &  2.04  & 3D : 1.0   \\  
mp-1209978  & V  & \ce{NbSe2}  & -1.75  &  2.84  &  2.04  & 3D : 1.0   \\
mp-7443  & Cr  & \ce{NbSe2}  & -2.28  &  1.96  &  2.26  & 3D : 0.996   \\ 
mp-985289  & Cr  & \ce{NbSe2}  & -2.08  &   2.6  &  2.04  & 3D : 1.0   \\ 
mp-1193575  & Mn  & \ce{NbSe2}  & -2.09  &  2.63  &  2.07  & 3D : 0.992   \\
mp-1190037  & Fe  & \ce{NbSe2}  & -1.49  &  2.76  &  2.04  & 3D : 0.986   \\
mp-1193638  & Fe  & \ce{NbSe2}  &  -1.2  &  2.73  &  2.07  & 3D : 0.987   \\
mp-1186227  & Co  & \ce{NbSe2}  & -1.03  &  2.83  &  2.11  & 3D : 0.963   \\
mp-1193368  & Co  & \ce{NbSe2}  & -0.751  &  2.68  &  2.07  & 3D : 0.962   \\
mp-20279  & In  & \ce{NbSe2}  & -0.917  &   2.8  &  2.07  & 3D : 0.534   \\
mp-1018116  & In  & \ce{NbSe2}  & -1.38  &  2.92  &  2.05  & 02D : 0.712   \\
\hline 
mp-755664  & Li  & \ce{TaS2}  & -1.48  &  3.19  &  2.21  & 3D : 0.754   \\ 
mp-1206881  & Li  & \ce{TaS2}  & -1.85  &  2.78  &  2.26  & 3D : 0.773   \\
mp-1222716  & Li  & \ce{TaS2}  & -1.63  &  2.98  &  2.26  & 2D : 0.683   \\
mp-1221434  & Na  & \ce{TaS2}  & -1.48  &  3.34  &  2.26  & 3D : 0.917   \\ 
mp-1223710  & K  & \ce{TaS2}  & -1.79  &   3.3  &  2.26  & 3D : 0.991   \\ 
mp-1190092  & V  & \ce{TaS2}  & -1.28  &  3.24  &  2.26  & 3D : 1.0   \\ 
mp-1189011  & Cr  & \ce{TaS2}  & -1.44  &  3.11  &  2.37  & 3D : 1.0   \\
mp-3581  & Mn  & \ce{TaS2}  & -1.63  &   2.8  &  2.22  & 3D : 0.989   \\   
mp-1208432  & Mn  & \ce{TaS2}  & -1.79  &  2.96  &  2.22  & 3D : 1.0   \\   
mp-1218077  & Ni  & \ce{TaS2}  & -1.48  &  2.15  &  2.22  & 2D : 0.968   \\ 
mp-1218136  & Ni  & \ce{TaS2}  & -1.46  &   2.4  &  2.26  & 2D : 0.919   \\ 
mp-1219081  & Ni, Cr  & \ce{TaS2}  & -1.51  &  2.08  &  2.22  & 2D : 0.966   \\
mp-554416  & Fe  & \ce{TaS2}  & -0.752  &  2.88  &  2.21  & 3D : 0.987   \\ 
mp-1218020  & Fe  & \ce{TaS2}  & -2.24  &  2.59  &  2.22  & 2D : 0.959   \\ 
mp-1218051  & Mo  & \ce{TaS2}  &  -1.7  &  2.21  &  2.26  & 2D : 0.965   \\
mp-22332  & In  & \ce{TaS2}  & -1.05  &  3.12  &  2.28  & 02D : 0.528   \\
mp-1101055  & In  & \ce{TaS2}  & -1.47  &  2.92  &  2.24  & 02D : 0.551   \\ 
mp-1218110  & In  & \ce{TaS2}  & -1.15  &  2.86  &  2.24  & 02D : 0.591   \\
mp-1218117  & In  & \ce{TaS2}  & -1.27  &  3.03  &  2.22  & 02D : 0.591   \\
mp-1221424  & Na  & \ce{TaSe2}  &  -1.8  &  2.35  &  2.44  & 3D : 0.946   \\
mp-1223811  & K  & \ce{TaSe2}  & -2.93  &  1.72  &  2.49  & 2D : 0.981   \\  
mp-1188354  & V  & \ce{TaSe2}  & -2.65  &  1.86  &  2.44  & 3D : 1.0   \\ 
mp-1187280  & Cr  & \ce{TaSe2}  & -2.73  &  1.77  &  2.44  & 3D : 1.0   \\ 
mp-1208377  & Cr  & \ce{TaSe2}  & -2.35  &  2.19  &  2.47  & 3D : 1.0   \\  
mp-999136  & In  & \ce{TaSe2}  & -1.09  &  2.72  &  2.45  & 02D : 0.507   \\ 
mp-1208590  & In  & \ce{TaSe2}  & -1.43  &  2.56  &  2.48  & 02D : 0.681   \\
mp-1218192  & In  & \ce{TaSe2}  & -1.42  &  2.51  &  2.48  & 02D : 0.702   \\

\hline 
mp-30533  & K  & \ce{Pt2S3}  & -0.836  &  1.71  &  1.62  & 3D : 0.933   \\
mp-28987  & Na  & \ce{Pt2Se3}  & -1.25  &  1.67  &  2.13  & 3D : 0.731   \\
mp-14796  & K  & \ce{Pt2Se3}  & -1.02  &  1.68  &  2.14  & 3D : 0.83   \\ 
mp-14797  & Rb  & \ce{Pt2Se3}  & -0.939  &  1.68  &  2.13  & 3D : 0.977   \\ 
mp-573316  & Cs  & \ce{Pt2Se3}  & -0.96  &  1.68  &  2.14  & 3D : 1.0   \\ 
\hline
mp-1247173  & Mg, Al  & \ce{Mn3S8}  & -1.31  &  1.15  & 0.854  & 3D : 0.792   \\
mp-1384478  & Mn, Zn  & \ce{Mn3S8}  & -1.08  & 0.902  & 0.854  & 3D : 0.714   \\ 
mp-1410942  & Mn, Zn  & \ce{Mn3S8}  & -0.976  & 0.789  & 0.852  & 3D : 0.6   \\ 
mp-1443978  & Mn, Mg  & \ce{Mn3S8}  & -0.834  & 0.965  & 0.855  & 3D : 0.94   \\
\hline 
mp-17228  & K  & \ce{Ni3S4}  & -1.28  & 0.934  &  1.23  & 3D : 0.799   \\ 
mp-672177  & K  & \ce{Ni3S4}  & -1.17  & 0.885  &  1.23  & 3D : 0.86   \\ 
mp-1079718  & Rb  & \ce{Ni3S4}  &  -1.2  &  1.05  &  1.23  & 3D : 0.915   \\
mp-28486  & Cs  & \ce{Ni3S4}  & -1.12  &  1.14  &  1.23  & 3D : 0.994   \\  
mp-1080141  & Cs  & \ce{Ni3S4}  &  -1.2  &   1.1  &  1.23  & 3D : 1.0   \\ 
mp-9910  & K  & \ce{Pd3S4}  & -1.11  &  1.51  &  1.11  & 3D : 0.823   \\ 
mp-11695  & Rb  & \ce{Pd3S4}  & -1.11  &  1.48  &  1.08  & 3D : 0.918   \\ 
mp-663190  & Rb  & \ce{Pd3S4}  & -1.06  &  1.51  &   1.2  & 3D : 0.946   \\ 
mp-1205388  & Rb  & \ce{Pd3S4}  & -1.11  &  1.49  &  1.09  & 3D : 0.924   \\
mp-510268  & Cs  & \ce{Pd3S4}  & -1.12  &  1.48  &  1.09  & 3D : 0.995   \\ 
mp-14339  & K  & \ce{Pd3Se4}  & -1.01  &  1.26  &   1.3  & 3D : 0.739   \\ 
mp-683041  & K  & \ce{Pd3Se4}  & -0.917  &  1.34  &  1.53  & 3D : 0.717   \\
mp-14340  & Rb  & \ce{Pd3Se4}  & -0.929  &  1.27  &  1.32  & 3D : 0.852   \\
mp-683059  & Rb  & \ce{Pd3Se4}  & -0.817  &  1.31  &  1.52  & 3D : 0.858   \\
mp-11694  & Cs  & \ce{Pd3Se4}  & -0.902  &   1.3  &   1.3  & 3D : 0.967   \\ 
mp-4030  & Rb  & \ce{Pt3S4}  & -1.09  &  1.41  &  1.47  & 3D : 0.907   \\ 
mp-663224  & Rb  & \ce{Pt3S4}  & -1.03  &   1.4  &  1.59  & 3D : 0.918   \\
mp-13992  & Cs  & \ce{Pt3S4}  & -1.08  &  1.43  &  1.46  & 3D : 0.995   \\  
mp-14338  & Cs  & \ce{Pt3Se4}  & -0.937  &  1.27  &  1.88  & 3D : 0.968   \\
\hline 
mp-1079772  & Ca  & \ce{InP}  &  -3.2  & 0.816  & 0.779  & 3D : 0.894   \\ 
mp-1078973  & Sr  & \ce{InP}  & -3.41  & 0.843  & 0.793  & 3D : 0.866   \\ 
mp-1078002  & Mn  & \ce{GaSe2}  & -0.822  &  1.29  &  1.39  & 3D : 0.999   \\
mp-1199743  & Cs  & \ce{InTe2}  & -0.939  & 0.919  &  1.22  & 3D : 0.939   \\ 
\hline 
mp-505005  & Ce  & \ce{OsSi}  &  -4.9  &  1.56  &  1.81  & 3D : 1.0   \\ 
mp-754516  & Li  & \ce{NiP}  & -1.41  &  1.16  &  1.04  & 3D : 0.716   \\
mp-4767  & Ce  & \ce{OsSi}  & -4.26  &  1.45  &  1.81  & 3D : 1.0   \\ 
mp-1206349  & K  & \ce{CoP}  & -1.97  &  1.19  &  1.93  & 3D : 0.808   \\ 
mp-9473  & Ba  & \ce{NiP}  & -3.02  &  1.01  &  1.05  & 3D : 0.903   \\ 
mp-1206920  & K  & \ce{IrP}  &  -1.8  &  1.22  &  1.56  & 3D : 0.667   \\ 
mp-1206529  & Rb  & \ce{IrP}  & -1.88  &  1.31  &  1.55  & 3D : 0.786   \\ 
mp-4815  & Pr  & \ce{RhSi}  & -4.42  &  2.38  &  2.56  & 3D : 1.0   \\ 
mp-10698  & Ba  & \ce{RhGe}  & -3.38  &  1.78  &  2.02  & 3D : 0.932   \\ 
mp-1207113  & Rb  & \ce{RhP}  & -2.06  &   1.5  &  1.59  & 3D : 0.785   \\ 
mp-567408  & Sm  & \ce{OsSi}  & -4.19  &  1.29  &  1.81  & 3D : 1.0   \\ 
mp-1206405  & Cs  & \ce{RhP}  & -2.18  &  1.51  &  1.59  & 3D : 0.928   \\ 
mp-11169  & Ba  & \ce{IrP}  & -3.93  &  1.09  &  1.56  & 3D : 0.878   \\ 
mp-571586  & Nd  & \ce{OsSi}  & -4.25  &  1.27  &  1.81  & 3D : 1.0   \\ 
mp-567203  & La  & \ce{OsSi}  & -4.31  &  1.32  &  1.81  & 3D : 1.0   \\  
mp-1192652  & La, Al  & \ce{OsB}  & -3.41  &  1.52  &  5.05  & 3D : 1.0   \\
mp-978853  & Sr  & \ce{IrGe}  & -3.21  &  1.23  &  1.77  & 3D : 0.985   \\ 
mp-5936  & La  & \ce{RhSi}  &  -4.4  &  2.37  &  2.56  & 3D : 1.0   \\ 
mp-12098  & K  & \ce{RhP}  & -1.99  &  1.44  &  1.59  & 3D : 0.68   \\ 
mp-3585  & La  & \ce{IrSi}  & -3.94  &  2.25  &  1.84  & 3D : 1.0   \\ 
mp-5852  & Pr  & \ce{OsSi}  &  -4.4  &  1.29  &  1.81  & 3D : 1.0   \\ 
mp-12073  & Ba  & \ce{IrB}  & -3.74  &  1.01  & 0.866  & 3D : 1.0   \\ 
mp-1206424  & Sr  & \ce{PtSi}  & -3.69  &  1.86  &  1.14  & 3D : 0.997   \\ 
mp-21849  & Eu  & \ce{IrSi}  & -3.43  &  2.15  &  1.84  & 3D : 1.0   \\ 
mp-1207365  & Cs  & \ce{IrP}  & -2.02  &  1.26  &  1.56  & 3D : 0.932   \\ 
mp-21383  & Eu  & \ce{RhSi}  & -3.77  &  2.29  &  2.56  & 3D : 1.0   \\ 
mp-10697  & Sr  & \ce{RhGe}  & -3.47  &  1.69  &  2.02  & 3D : 0.987   \\ 
mp-8581  & Sr  & \ce{RhP}  &  -3.9  &  1.29  &  1.61  & 3D : 0.849   \\ 
mp-1206941  & Rb  & \ce{CoP}  & -2.24  &  1.23  &  1.93  & 3D : 0.889   \\ 
mp-8583  & Ba  & \ce{RhP}  & -4.12  &  1.37  &  1.62  & 3D : 0.89   \\  
mp-8982  & Ca  & \ce{PtSi}  & -3.49  &  1.88  &  2.07  & 3D : 1.0   \\  
mp-20615  & Eu  & \ce{PdGe}  & -3.55  &  1.46  &  1.23  & 3D : 1.0   \\
mp-978253  & Ce  & \ce{RhSi}  & -4.79  &  2.34  &  2.53  & 3D : 0.988   \\ 
mp-1193516  & Eu  & \ce{PdSn}  & -3.89  &  1.41  &  1.05  & 3D : 1.0   \\ 
mp-504772  & Y  & \ce{RhSi}  & -4.83  &  2.32  &  2.53  & 3D : 1.0   \\ 
mp-1211651  & La  & \ce{RhGe}  &  -4.9  &  1.55  &  1.43  & 3D : 1.0   \\ 
mp-1208977  & Sm  & \ce{RhSi}  & -4.55  &  2.21  &  2.53  & 3D : 1.0   \\ 
mp-1208966  & Sm  & \ce{PdSi}  & -4.45  &  2.02  &  1.41  & 3D : 0.965   \\ 
mp-627355  & Ce  & \ce{PtGe}  &  -3.9  &  1.98  &   1.1  & 3D : 1.0   \\
\hline 
mp-1215178  & Zr  & \ce{TiB4}  & -2.06  & 0.867  &  1.26  & 3D : 1.0   \\ 
mp-1224263  & Hf  & \ce{TiB4}  & -2.04  & 0.787  &  1.25  & 3D : 1.0   \\ 
mp-1215211  & Zr  & \ce{NbB4}  & -2.44  &  1.56  &  2.54  & 3D : 1.0   \\  
mp-1224328  & Hf  & \ce{NbB4}  & -2.59  &  1.48  &  2.55  & 3D : 1.0   \\ 
mp-1215209  & Zr  & \ce{TaB4}  & -2.72  &   1.5  &   2.5  & 3D : 1.0   \\ 
mp-1224283  & Hf  & \ce{TaB4}  & -2.65  &  1.46  &  2.52  & 3D : 1.0   \\ 
mp-1220697  & Al  & \ce{Nb2B6}  &  -4.9  & 0.772  &  1.45  & 02D : 0.532   \\
mp-996161  & Al  & \ce{Nb3C2}  &  -3.2  &  1.13  &  1.36  & 3D : 0.997   \\
mp-569568  & Al  & \ce{Ta3C2}  & -2.65  &  1.11  &  1.06  & 3D : 1.0   \\ 
\hline 
mp-1216678  & Ti  & \ce{MoP2}  & -1.07  &  1.65  &  1.15  & 3D : 1.0   \\  
mp-1216676  & Ti  & \ce{WP2}  & -0.924  & 0.951  &  0.78  & 3D : 1.0   \\  
mp-1104130  & Ba  & \ce{AuP2}  & -2.55  & 0.824  & 0.969  & 3D : 0.948   \\
mp-1246556  & Sr  & \ce{VN2}  & -3.57  &  1.66  & 0.814  & 3D : 1.0   \\  
mp-1029711  & Na  & \ce{VN2}  & -1.22  &  1.74  & 0.815  & 3D : 1.0   \\ 
mp-1246136  & Ba  & \ce{VN2}  &  -4.6  &  1.28  & 0.816  & 3D : 1.0   \\ 
mp-1245629  & Ba  & \ce{VN2}  & -3.85  &  1.85  & 0.816  & 3D : 1.0   \\ 
mp-752942  & Li  & \ce{TiO2}  &  -1.1  &  1.83  &  2.62  & 3D : 0.993   \\
mp-780233  & Li  & \ce{TiO2}  & -1.12  &  1.01  &  2.63  & 3D : 1.0   \\ 
mp-1387047  & Ca  & \ce{TiO2}  & -1.64  &  1.52  &  2.64  & 3D : 1.0   \\  
mp-760000  & Li, V  & \ce{TiO2}  &  -1.2  & 0.774  &  2.68  & 3D : 0.992   \\
mp-757308  & Na  & \ce{TiO2}  & -1.46  &  1.27  &  2.64  & 3D : 1.0   \\ 
mp-1101470  & Na  & \ce{TiO2}  & -1.72  &  1.58  &  2.79  & 3D : 1.0   \\ 
mp-768342  & Sr  & \ce{NbO2}  & -2.86  &  2.64  & 0.876  & 3D : 1.0   \\ 
mp-867955  & Na  & \ce{NbO2}  & -0.854  &  2.31  & 0.752  & 3D : 0.989   \\
mp-29792  & Ca  & \ce{NbO2}  & -1.99  &  2.78  &  0.93  & 3D : 1.0   \\ 
mp-31908  & Mn  & \ce{Nb6O11}  & -2.12  &  2.52  &  1.76  & 3D : 0.943   \\
mp-760837  & Rb  & \ce{Nb10O17}  & -1.02  &  2.51  &  2.46  & 3D : 1.0   \\
mp-8236  & Ba  & \ce{PdP}  & -3.12  & 0.931  &  1.38  & 3D : 0.993   \\ 
mp-1227860  & Ba  & \ce{PdP}  & -3.39  &  1.39  &  1.24  & 3D : 1.0   \\ 
mp-28339  & Ca  & \ce{PtP}  & -2.33  &  1.85  &   3.1  & 3D : 0.919   \\  
mp-1218299  & Eu, Sr  & \ce{PtP}  & -2.32  &  1.62  & 0.817  & 3D : 1.0   \\
mp-685613  & Ba  & \ce{Pd2P}  &  -4.3  &  1.09  &  1.31  & 3D : 1.0   \\ 
mp-504701  & Cs  & \ce{TiF4}  & -1.45  &   2.5  &  1.04  & 3D : 1.0   \\  
mp-21639  & K  & \ce{TiF4}  & -0.949  &  2.52  &  1.13  & 3D : 0.965   \\ 
mp-1120745  & Na  & \ce{TiF4}  & -0.947  &  2.52  &  1.07  & 3D : 1.0   \\
mp-27264  & Na  & \ce{TiF4}  & -0.996  &  2.48  &  1.16  & 3D : 1.0   \\ 
mp-754779  & Li  & \ce{NbS3}  & -0.959  &  2.86  &  1.46  & 3D : 0.638   \\ 
mp-769050  & Li  & \ce{NbS3}  &  -0.9  &  2.62  &  1.47  & 3D : 0.636   \\ 
mp-2492941  & Li  & \ce{NbS3}  & -0.907  &  2.62  &  1.48  & 3D : 0.616   \\ 
mp-570823  & Ba  & \ce{BSe3}  & -1.85  &   1.0  & 0.893  & 2D : 0.509   \\
mp-30105  & Ba  & \ce{B4Se13}  & -1.72  & 0.929  &   1.3  & 3D : 0.986   \\ 
mp-28463  & Li  & \ce{Nb3Cl8}  & -0.815  &  2.35  &  1.65  & 3D : 0.935   \\  
\hline
\caption{Best candidates binary etched layers. The values of the highest etched energy (HEE), lowest non-etched energy (LNE) and lowest defect formation energy (LDFE) are given in eV.}
\end{longtable}

\newpage

\begin{table*}[h]
\begin{tabular}{lcccccc}
\hline
MP ID      & Etched & Layer     & HEE (eV) & LNE (eV) & LDFE (eV) & Initial Dim. \\ 
\hline
mp-1226508  & Ce  & \ce{OsRuSi2}  & -4.03  & 0.534  & 0.692  & 3D : 1.0   \\ 
mp-1220305  & Nd  & \ce{RuRhSi2}  & -3.79  & 0.502  & 0.597  & 3D : 1.0   \\ 
mp-1225172  & Eu  & \ce{Co2GeSi}  & -2.98  & 0.594  & 0.731  & 3D : 1.0   \\ 
mp-7129  & Rb  & \ce{TiCu2S4}  &  -1.2  & 0.795  &  1.28  & 3D : 0.968   \\ 
mp-10489  & Cs  & \ce{TiCu2Se4}  &  -1.2  & 0.585  &   1.5  & 3D : 0.978   \\ 
mp-10488  & Cs  & \ce{TiAg2S4}  & -0.992  & 0.541  &  1.18  & 3D : 0.948   \\ 
\hline
mp-1220884  & Na  & \ce{TiVS4}  & -0.984  &  1.59  &  1.44  & 3D : 0.899   \\ 
mp-1222735  & Li  & \ce{CrVS4}  & -1.22  & 0.951  &  1.22  & 3D : 0.644   \\ 
mp-1223464  & K  & \ce{CrSnS4}  & -0.841  &  1.54  &  1.13  & 3D : 0.923   \\ 
mp-2222797  & Mg  & \ce{Nb2PdS6}  & -3.05  &  1.65  &  1.16  & 3D : 0.923   \\ 
\hline
mp-1228980  & Al  & \ce{CuGeSe4}  & -2.06  & 0.889  & 0.674  & 3D : 0.992   \\ 
mp-1228963  & Al  & \ce{AgSnSe4}  & -1.84  & 0.741  & 0.609  & 3D : 0.992   \\ 
\hline
mp-1220518  & Ni  & \ce{Nb2CS2}  & -0.624  &  1.37  &   1.4  & 3D : 0.966   \\ 
mp-1218138  & Ni  & \ce{Ta2CS2}  & -0.926  &  1.62  &  1.71  & 3D : 0.901   \\ 
\hline
mp-1195350  & Cs  & \ce{CdInTe3}  & -0.872  & 0.723  &  1.38  & 3D : 0.892   \\ 
mp-541407  & Rb  & \ce{CuSnS3}  & -0.655  &  1.18  &  1.11  & 3D : 0.809   \\
mp-21713  & K  & \ce{CuIn3Se6}  & -0.738  & 0.697  &  1.14  & 3D : 0.791   \\ 
mp-680403  & K  & \ce{AgIn3Se6}  & -0.649  & 0.662  &  1.13  & 3D : 0.956   \\  
\hline
mp-6376  & K  & \ce{VCu2S4}  & -0.533  & 0.967  &  1.47  & 3D : 0.646   \\ 
mp-10091  & K  & \ce{VCu2Se4}  & -0.643  & 0.792  &  1.07  & 3D : 0.679   \\ 
mp-1106233  & K  & \ce{NbCu2S4}  & -0.629  & 0.691  &  1.17  & 3D : 0.56   \\
mp-6599  & K  & \ce{NbCu2Se4}  &  -0.7  & 0.656  & 0.915  & 3D : 0.642   \\ 
mp-505321  & Cs  & \ce{NbCu2Te4}  & -0.613  & 0.685  & 0.813  & 3D : 0.948   \\ 
mp-6013  & K  & \ce{TaCu2Se4}  & -0.687  & 0.668  &  1.04  & 3D : 0.623   \\ 
mp-505322  & Cs  & \ce{TaCu2Te4}  & -0.723  & 0.599  &  0.78  & 3D : 0.939   \\ 
mp-571288  & K  & \ce{TaAg2Se4}  & -0.716  &  0.53  & 0.855  & 3D : 0.474   \\ 
\hline
mp-1192531  & K  & \ce{Cu2SbS3}  & -0.739  & 0.575  &  1.05  & 3D : 0.94   \\ 
mp-1194436  & K  & \ce{Cu2BiS3}  & -0.735  & 0.564  &  1.15  & 3D : 0.95   \\ 
mp-1176770  & Li  & \ce{CrP2S7}  & -1.34  &  1.02  & 0.836  & 02D : 0.787   \\ 
mp-510569  & Cs  & \ce{CeCuS3}  & -0.712  & 0.879  &   1.3  & 3D : 0.946   \\ 
mp-505171  & Na  & \ce{TiCuS3}  & -0.881  & 0.703  &  1.18  & 3D : 0.73   \\ 
mp-1220939  & Na  & \ce{TiNCl}  & -0.798  &   2.7  &  2.45  & 3D : 0.594   \\ 
mp-679669  & Na  & \ce{Zr2N2ClS}  & -1.32  &  3.11  &  1.09  & 3D : 0.947   \\ 
mp-1220698  & Na  & \ce{Zr2N2ClS}  & -1.12  &  2.73  & 0.774  & 3D : 0.749   \\ 
\hline
\end{tabular}
\caption{Best candidates ternary layers. The values of the highest etched energy (HEE), lowest non-etched energy (LNE) and lowest defect formation energy (LDFE) are given in eV.}
\end{table*}

\newpage

\section{Chemical potentials}

\begin{figure*}[h]
    \centering
    \includegraphics[width=\linewidth]{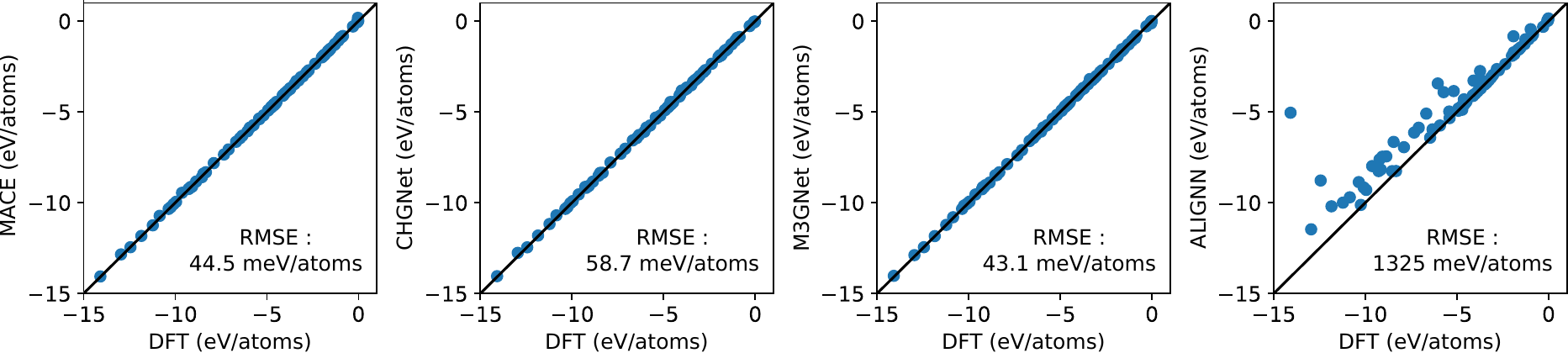}
    \caption{Comparison of the atomic formation energy from DFT calculations with different UMLIPs. Reference DFT values are taken from the Materials Project database.}
    \label{fig:Benchmark_potentials}
\end{figure*}

Two sets of chemical potentials are used in the main text: the elemental ones $\mu^E$ and the chemical potential in HF $\mu^\text{HF}$. The former is obtained from the structures with lowest formation energies available in the MP database. Note that the potentials $\mu^\text{E}$ are then computed using all four UMLIPs. A comparison between UMLIPs and DFT is presented in Fig.\ \ref{fig:Benchmark_potentials}. While MACE, CHGNet and M3GNet lead to extremely accurate atomic energies, ALIGNN fails for some elements, resulting in a much higher RMSE. This could in part explain the less accurate predictions of the vacancy formation energies presented in Fig.\ 1 of the main text.

For chemical potentials in HF, we used the ones previously used in Ref.\ \citenum{Bjork2024}. Additionally, we also included potentials of lanthanides using the same method. All chemical potentials are presented in Tables \ref{tab:chem_pot_1} and \ref{tab:chem_pot_2} below. Note that for simplicity, the table contains the value $\mu^\text{HF}-\mu^\text{E}$ rather than $\mu^\text{HF}$.

\newpage

\begin{table*}[h]
\begin{tabular}{cccc}
\hline
\hspace{5pt} Element \hspace{5pt} & \hspace{5pt} $\mu^\text{E}$ from MP \hspace{5pt} & \hspace{5pt} $\mu^\text{E}$ from MACE \hspace{5pt} & \hspace{5pt} $\mu^\text{HF}-\mu^\text{E}$ \hspace{5pt} \\
\hline
H     &     -3.39     &     -3.30     &     0.00     \\

Li     &     -1.90     &     -1.91     &     -3.22     \\ 
Be     &     -3.74     &     -3.75     &     -4.24     \\ 
B     &     -6.68     &     -6.65     &     0.00     \\ 
C     &     -9.23     &     -9.21     &     0.00     \\ 
N     &     -8.33     &     -8.32     &     0.00     \\ 
O     &     -4.92     &     -4.90     &     -2.46     \\ 
F     &     -1.87     &     -1.87     &     -3.00     \\ 

Na     &     -1.31     &     -1.31     &     -2.89     \\ 
Mg     &     -1.59     &     -1.61     &     -4.89     \\ 
Al     &     -3.74     &     -3.71     &     -5.84     \\ 
Si     &     -5.42     &     -5.37     &     0.00     \\ 
P     &     -5.41     &     -5.40     &     0.00     \\ 
S     &     -4.14     &     -4.11     &     0.00     \\ 
Cl     &     -1.84     &     -1.84     &     0.00     \\ 

K     &     -1.09     &     -1.07     &     -3.12     \\ 
Ca     &     -2.00     &     -2.00     &     -5.92     \\ 
Sc     &     -6.33     &     -6.31     &     -6.26     \\ 
Ti     &     -7.90     &     -7.82     &     -3.80     \\ 
V     &     -9.08     &     -9.12     &     -2.69     \\ 
Cr     &     -9.63     &     -9.45     &     -2.32     \\ 
Mn     &     -9.16     &     -9.14     &     -2.54     \\ 
Fe     &     -8.46     &     -8.40     &     -1.13     \\ 
Co     &     -7.09     &     -7.07     &     -0.74     \\ 
Ni     &     -5.73     &     -5.73     &     -0.65     \\ 
Cu     &     -4.10     &     -4.09     &     0.00     \\ 
Zn     &     -1.26     &     -1.25     &     -1.70     \\ 
Ga     &     -3.03     &     -3.04     &     -2.11     \\ 
Ge     &     -4.62     &     -4.60     &     0.00     \\ 
Se     &     -3.49     &     -3.47     &     0.00     \\ 
Br     &     -1.64     &     -1.62     &     -0.07     \\ 
\hline
\end{tabular}
\caption{\label{tab:chem_pot_1} Chemical potentials for elements from H to Cl. Both the elemental potentials and the potentials in HF are presented.}
\end{table*}

\newpage

\begin{table*}[h]
\begin{tabular}{cccc}
\hline
\hspace{5pt} Element \hspace{5pt} & \hspace{5pt} $\mu^\text{E}$ from MP \hspace{5pt} & \hspace{5pt} $\mu^\text{E}$ from MACE \hspace{5pt} & \hspace{5pt} $\mu^\text{HF}-\mu^\text{E}$ \hspace{5pt} \\
\hline
Rb     &     -0.98     &     -0.91     &     -3.12     \\ 
Sr     &     -1.68     &     -1.68     &     -5.98     \\ 
Y     &     -6.46     &     -6.45     &     -7.37     \\ 
Zr     &     -8.54     &     -8.59     &     -5.96     \\ 
Nb     &     -10.09     &     -10.08     &     0.00     \\ 
Mo     &     -10.85     &     -10.72     &     0.00     \\ 
Ru     &     -9.28     &     -9.25     &     -1.38     \\ 
Rh     &     -7.34     &     -7.35     &     0.00     \\ 
Pd     &     -5.18     &     -5.19     &     0.00     \\ 
Ag     &     -2.83     &     -2.82     &     0.00     \\ 
Cd     &     -0.91     &     -0.88     &     -0.98     \\ 
In     &     -2.72     &     -2.71     &     -1.39     \\ 
Sn     &     -4.01     &     -3.97     &     -0.65     \\ 
Sb     &     -4.12     &     -4.07     &     0.00     \\ 
Te     &     -3.14     &     -3.09     &     0.00     \\ 
I     &     -1.52     &     -1.52     &     -0.00     \\

Cs     &     -0.85     &     -0.82     &     -3.20     \\ 
Ba     &     -1.92     &     -1.91     &     -5.99     \\ 
La     &     -4.93     &     -4.92     &     -7.26     \\ 
Ce     &     -5.93     &     -5.87     &     -7.14     \\
Pr     &     -4.77     &     -4.76     &     -7.22     \\ 
Nd     &     -4.76     &     -4.74     &     -7.14     \\ 
Sm     &     -4.71     &     -4.70     &     -7.09     \\ 
Eu     &     -10.25     &     -10.25     &     -6.13     \\ 
Gd     &     -14.09     &     -14.06     &     -7.03     \\ 
Tb     &     -4.62     &     -4.61     &     -6.93     \\ 
Dy     &     -4.59     &     -4.59     &     -7.07     \\ 
Ho     &     -4.58     &     -4.56     &     -7.16     \\ 
Er     &     -4.57     &     -4.56     &     -7.11     \\ 
Tm     &     -4.47     &     -4.45     &     -7.04     \\
Lu     &     -4.51     &     -4.52     &     -6.83     \\ 
Hf     &     -9.96     &     -9.95     &     -5.93     \\ 
Ta     &     -11.85     &     -11.83     &     0.00     \\ 
W     &     -12.96     &     -12.85     &     0.00     \\ 
Re     &     -12.44     &     -12.45     &     -0.52     \\ 
Os     &     -11.22     &     -11.25     &     0.00     \\ 
Ir     &     -8.86     &     -8.84     &     0.00     \\ 
Pt     &     -6.05     &     -6.06     &     0.00     \\ 
Au     &     -3.27     &     -3.27     &     0.00     \\ 
Hg     &     -0.30     &     -0.30     &     0.00     \\ 
Tl     &     -2.37     &     -2.35     &     -0.51     \\ 
Pb     &     -3.71     &     -3.69     &     -0.43     \\ 
Bi     &     -3.88     &     -3.84     &     0.00     \\ 
\hline
\end{tabular}
\caption{\label{tab:chem_pot_2} Chemical potentials for elements from K to Bi. Both the elemental potentials and the potentials in HF are presented.}
\end{table*}

\newpage

\section{UMLIP models in ASE}

For MACE, we use the pre-trained MACE-MP-0 model, with small model size and withouth dispersion. It is directly loaded from the mace python package using the following code:
\begin{lstlisting}[language=Python]
from mace.calculators import mace_mp
MACE = mace_mp(model="small", dispersion=False, default_dtype="float32", device='cuda')
\end{lstlisting}

\vspace{1cm}

For CHGNet, we use the default model from the chgnet python package, which can be loaded as:
\begin{lstlisting}[language=Python]
from chgnet.model.dynamics import CHGNetCalculator
CHGNet = CHGNetCalculator(use_device='cuda', on_isolated_atoms='ignore')
\end{lstlisting}

\vspace{1cm}

For M3GNet, we use the model trained on MP and it is loaded using the matgl package.
\begin{lstlisting}[language=Python]
import matgl
from matgl.ext.ase import PESCalculator
M3GNet = PESCalculator(matgl.load_model("M3GNet-MP-2021.2.8-PES"))
\end{lstlisting}

\vspace{1cm}

Similarly, the ALIGNN model trained on MP can be loaded using the alignn python package.
\begin{lstlisting}[language=Python]
from alignn.ff.ff import AlignnAtomwiseCalculator, mptraj_path
ALIGNN = AlignnAtomwiseCalculator(path=mptraj_path())
\end{lstlisting}

\bibliography{Vac_MACE}